\documentclass[a4paper,fleqn,usenatbib]{mnras}

\usepackage{newtxtext,newtxmath}

\usepackage[T1]{fontenc}
\usepackage{ae,aecompl}

\usepackage{graphicx}	
\usepackage{amsmath}	
\usepackage{amssymb}	

\newcommand{\diff}{\mathrm{d}}
\newcommand{\sh}{\mathrm{sh}}

\newcommand{\view}{\theta_\mathrm{v}}
\newcommand{\thj}{\theta_\mathrm{j}}

\newcommand{\emisd}{\epsilon' _{\nu '}}
\newcommand{\numd}{\nu'_\mathrm{m}}
\newcommand{\nucd}{\nu'_\mathrm{c}}
\newcommand{\me}{m_\mathrm{e}}
\newcommand{\mpr}{m_\mathrm{p}}
\newcommand{\qe}{q_\mathrm{e}}
\newcommand{\eB}{\varepsilon_\mathrm{B}}
\newcommand{\ee}{\varepsilon_\mathrm{e}}
\newcommand{\BM}{\mathrm{BM}}
\newcommand{\ST}{\mathrm{ST}}
\newcommand{\Ec}{E_\mathrm{c}}
\newcommand{\thc}{\theta_\mathrm{c}}

\newcommand{\fb}{f_\mathrm{b}}
\newcommand{\thb}{\theta_\mathrm{b}}
\newcommand{\ts}{t_\mathrm{s}}
\newcommand{\Rs}{R_\mathrm{s}}
\newcommand{\Tp}{T_\mathrm{p}}
\newcommand{\Tf}{T_\mathrm{f}}
\newcommand{\Fp}{F_{\nu,\mathrm{p}}}
\newcommand{\simpropto}{\mathrel{\vcenter{
			\offinterlineskip\halign{\hfil$##$\cr
				\propto\cr\noalign{\kern2pt}\sim\cr\noalign{\kern-2pt}}}}}

\title[Inverse reconstruction of jet structure]{Inverse reconstruction of jet structure from off-axis gamma-ray burst afterglows}

\author[Takahashi \& Ioka]{
Kazuya Takahashi$^{1}$\thanks{E-mail: kazuya.takahashi@yukawa.kyoto-u.ac.jp} 
and Kunihito Ioka$^{1}$
\\
$^{1}$Center for Gravitational Physics, Yukawa Institute for Theoretical Physics, Kyoto University, Kyoto, 606-8502, Japan
}

\date{Accepted XXX. Received YYY; in original form ZZZ}

\pubyear{2019}

\begin{document}
\label{firstpage}
\pagerange{\pageref{firstpage}--\pageref{lastpage}}
\maketitle

\begin{abstract}
The gravitational wave event GW170817 and the slowly-rising afterglows of short gamma-ray
burst GRB 170817A clearly suggest that the GRB jet has an angular structure. However the actual
jet structure remains unclear as different authors give different structures. We formulate a
novel method to inversely reconstruct the jet structure from off-axis GRB afterglows, without
assuming any functional form of the structure in contrast to the previous studies. 
The jet structure is uniquely determined from the rising part of a light curve for a given parameter set
by integrating an ordinary differential equation, which is derived from the standard theory
of GRB afterglows. Applying to GRB 170817A, we discover that a non-trivial hollow-cone
jet is consistent with the observed afterglows, as well as Gaussian and power-law jets
within errors, which implies the Blandford-Znajek mechanism or an ejecta-jet interaction. The
current observations only constrain the jet core, not in principle the outer jet structure around
the line of sight. More precise and high-cadence observations with our inversion method will
fix the jet structure, providing a clue to the jet formation and propagation.
\end{abstract}




\section{Introduction}

Formation and propagation of a relativistic jet is
one of the unresolved problems in astrophysics.
The problem is important for understanding the most violent phenomena in the universe,
such as Gamma-Ray Bursts (GRBs), active galactic nuclei, and microquasars.
Although the relativistic jet is thought to be launched by
the system of a compact star, such as a black hole or neutron star,
and accretion disc with magnetic fields,
the exact nature is not known
because various physical processes are involved,
such as general relativity,
Blandford-Znajek mechanism,
neutrino annihilation,
magnetic reconnection,
disc wind,
jet collimation,
baryon loading,
shock breakout, and so on.

The multi-messenger observations of the gravitational wave event GW170817
from a merger of two neutron stars \citep{PRL170817}
and the associated short GRB~170817A \citep{170817multi,170817FermiGBM,170817Integral}
give a new clue to the mystery of a relativistic jet.
It is now widely accepted that this event launched a relativistic jet
that successfully penetrated the ejecta from the neutron star merger \citep[e.g.,][]{superluminal,rapiddecline,Ghirlanda19,rapiddeclineHST,Troja19} and
the jet is off-axis to us, leading to the very faint GRB
by relativistic beaming as observed
\citep[e.g.,][]{170817gamma,IN18,IN19}.
The jet power should be similar to those in the other normal short GRBs,
otherwise the jet cannot penetrate the merger ejecta \citep{Nagakura14,Hamidani19}.

The afterglows show slowly-rising light curves in radio to X-ray.
This is not explained by a uniform jet (a.k.a. a top-hat jet) \citep{Mooley18} but by a jet that has an angular structure (a so-called structured jet), which
interacts with the ambient medium
and radiates synchrotron emission from electrons accelerated at the forward shock \citep[e.g.,][]{Lazzati18,Margutti18,Ghirlanda19,rapiddeclineHST,Troja19}.
The angular structure is also important for solving the spectral puzzles of GRB~170817A \citep{Kisaka+18,IN19,Matsumoto19a,Matsumoto19b}.
The jet structure obtained from the afterglow observations would constrain the formation and propagation of the jet.

However, various authors give various jet structures
(see fig.~1 in \citet{IN19} and fig.~1 in \citet{Ryan})
and the true jet structure is not settled yet.
Two types of jet structure are frequently discussed.
One is a Gaussian jet, where the profile of the isotropic equivalent energy of the jet is described by a Gaussian of the angle from the jet axis \citep{ZM02,Lyman18,Resmi18,Troja19,rapiddeclineHST}.
The other is a power-law jet, where the energy profile obeys a power law outside a core \citep{Meszaros98,Rossi02,ZM02,DAvanzo18,Ghirlanda19}.
These jet models contain model parameters that control the structure,
which are adjusted by fitting the synthesized light curves to the observed data.
Recently, \citet{Ryan} proposed a way to infer the jet structure from afterglow light curves, assuming a Gaussian or a power-law structure. The analytic formula for jet structure in \citet{GG18} also assumes a power-law structure beforehand.
In any case, the functional form of the jet structure is assumed at the beginning in these previous studies. In such a way, it would be challenging to determine the functional form itself.\footnote{It may be possible to determine the jet structure even if the functional form is assumed at the beginning, although it would be computationally and/or technically more challenging than our method. For example, one way is to try copious functional forms and find the best-fitting one. Another is to model the jet structure by a generic function with many parameters, such as a high order polynomial or a piecewise linear function with many segments, and tune the free parameters to fit the data.}

In this paper, we propose a novel method to determine the functional form of the jet structure itself. We consider an inverse problem, and inversely reconstruct the jet structure from off-axis GRB afterglows without assuming any functional form of the jet structure.
In our method, the energy distribution of the structured jet is automatically determined from the observed light curve by integrating an ordinary differential equation, which is formulated based on the standard theory of GRB afterglows. This is a sharp contrast to the previous methods mentioned above, which {\it a priori} assume a Gaussian or a power-law structure.
Furthermore, our method uniquely determines the jet structure for a given afterglow light curve and a given parameter set.
Applying the inversion method to GRB~170817A,
we find that a hollow-cone jet is also consistent with the observed afterglow light curves for the first time, as well as Gaussian and power-law jets.

The paper is organized as follows.
We formulate our inversion equation
after reviewing the synchrotron shock model of off-axis GRB afterglows
in Section~\ref{sec.method}.
The inversion formula is applied to the afterglow of GRB~170817A in Section~\ref{sec.results},
where we find a hollow-cone jet as well as Gaussian and power-law jets can explain the observed light curves.
In Section~\ref{sec.conclusion}, we briefly summarize this study and discuss remaining issues on the inversion method that will be investigated in a forthcoming paper.
Throughout the paper, we attach a prime to the quantities evaluated in the fluid rest frame.

\section{Method}\label{sec.method}
We inversely reconstruct the angular energy distribution of a GRB jet from the afterglow light curve for off-axis GRBs. The basic idea for the inverse reconstruction is that an off-axis observer sees more and more inner regions close to the jet axis as time goes, so that the later afterglow brings new information on more inner jet. In the early phase, before the afterglow shock is sufficiently decelerated, the observable region is limited to a small angle around the line of sight due to relativistic beaming effects. The observable region of the shock gradually expands as the jet decelerates and the relativistic beaming effects become weak. The newly observable region contributes to the afterglow flux, which reflects the energy contained in the region. Since the inner region is usually expected to be brighter than the outer one, the energy distribution can be inversely estimated from the rising part of the afterglow light curve before the jet break.

\subsection{Review of the synchrotron afterglow model} \label{sec.review}
We review here a theoretical model for calculating synchrotron emission of GRB afterglows. The basic equations reviewed in this subsection is the starting point of the inversion formula in Section~\ref{sec.Inv}. First, we explain the shock dynamics that is applied for relativistic and non-relativistic regimes. Then, the local synchrotron emissivity is described. Finally, the equation for the observed synchrotron emission is presented by incorporating these prescriptions. We note that the formulation is essentially the same as in \citet{Sari98, Eerten10}. 

We consider that a relativistic jet is adiabatically propagating in a stationary, cold, uniform ambient medium with a constant number density $n_0$. The jet is assumed to be axisymmetric and has an angle-dependent energy distribution. We also assume that each jet segment spherically expands as if it is a portion of an isotropic blast wave that has the same isotropic equivalent energy. This assumption holds well for a relativistic jet unless it is decelerated sufficiently below a local sound speed and each segment interacts with each other \citep{KG03,ZM09,EM12}. Then, the dynamics of each shock segment would be well described by a self-similar solution of \citet{BM}. As the shock decelerated to a non-relativistic speed by sweeping the ambient material, the shock dynamics is better described by the Sedov-Taylor self-similar solution \citep{Sedov,Taylor} rather than the Blandford-McKee solution. In order to smoothly connect the relativistic and non-relativistic regimes, we describe shock propagation as a hybrid of these two self-similar solutions as follows \citep{Eerten10}:
\begin{align}
\label{eq.BM-STsh}
\Gamma_\sh^2 \beta_\sh^2 &= C_\BM^2 t^{-3} + C_\ST^2t^{-6/5},\\
\label{eq.BM-ST}
\Gamma^2 \beta^2 &= \frac{1}{2}C_\BM^2 t^{-3} + \frac{9}{16}C_\ST^2t^{-6/5},
\end{align}
where $\beta_\sh$ and $\beta$ are the speeds of the shock wave and shocked fluid normalized by the speed of light $c$, respectively, and $\Gamma_\sh = 1/\sqrt{1 - \beta_\sh^2}$ and $\Gamma = 1/\sqrt{1 - \beta^2}$ are the corresponding Lorentz factors. $t$ denotes the elapsed laboratory time since the explosion. The coefficients $C_\BM$ and $C_\ST$ are given by
\begin{align}
\label{eq.C_BM}
C_\BM &= \sqrt{\frac{17E}{8\pi n_0\mpr c^5}},\\
\label{eq.C_ST}
C_\ST &= \frac{2}{5}\cdot 1.15 \left(\frac{E}{n_0 \mpr c^5}\right)^{1/5},
\end{align}
where $E=E(\theta)$ is the isotropic equivalent energy, which is defined for each unit solid angle of the structured jet, and $\mpr$ stands for the proton mass. The factors $1/2$ and $9/16$ in Equation~(\ref{eq.BM-ST}) come from the strong shock jump conditions in the relativistic (with the ratio of the specific heats $\hat{\gamma}=4/3$) and non-relativistic (with $\hat{\gamma}=5/3$) limits, respectively. The numerical factor $1.15$ in Equation~(\ref{eq.C_ST}) comes from the energy conservation. We note that Equations~(\ref{eq.BM-STsh}) and (\ref{eq.BM-ST}) are reduced to the Blandford-McKee solution by formally putting $\beta = \beta_\sh = 1$ and neglecting the second term. The radius of each shock segment at a given laboratory time $t$ is given by integrating the shock speed:
\begin{equation}
\label{eq.R}
R  = \int _0^t c\beta_\sh\diff t.
\end{equation}

The local synchrotron emission at the fluid rest frame is evaluated based on the standard model of GRB afterglows \citep{Sari98}, where microscopic physics such as the amplification of magnetic fields and particle acceleration through the shock wave is modelled by introducing phenomenological parameters, $\eB$ and $\ee$, respectively. In this model, the non-thermal electrons have an isotropic energy distribution described by a simple power law with an index $p$ in the fluid rest frame. The magnetic field in the shock downstream is assumed to be well tangled and, hence, the synchrotron emission is isotropic in the fluid rest frame. We neglect synchrotron-self absorption henceforth, because it is not relevant for our arguments. Then, the spectrum is well approximated by a broken power law bent at the synchrotron characteristic frequency $\numd$ and cooling frequency $\nucd$. In the case of slow cooling ($\numd < \nucd$), the energy radiated by synchrotron emission per unit volume per unit time per unit frequency $\emisd$ is given by
\begin{equation}
\label{eq.slowcooling}
\emisd = \epsilon '_{\nu',\mathrm{p}}\times \left\{
\begin{array}{ll}
\displaystyle \left(\frac{\nu'}{\numd}\right)^{1/3} & (\nu' < \numd)\\
\displaystyle \left(\frac{\nu'}{\numd}\right)^{-(p-1)/2} & (\numd \le \nu' < \nucd) \\
\displaystyle \left(\frac{\nucd}{\numd}\right)^{-(p-1)/2}\left(\frac{\nu'}{\nucd}\right)^{-p/2} & (\nucd \le \nu')
\end{array} \right.. 
\end{equation}
The peak emissivity
is given by \citep{Granot99,Eerten10}\footnote{The numerical coefficients in Equations~(\ref{eq.emisdpeak}), (\ref{eq.numd}), and (\ref{eq.nucd}) are different from those in \citet{Sari98} but taken from \citet{Granot99}, who more accurately fitted the broken power-law spectrum to the exact one \citep{RL}. We confirmed the numerical factor $0.88$ in Equation~(\ref{eq.emisdpeak}) is also valid for $p=2.17$, while the factor was originally introduced for $p=2.5$ to adjust the spectrum.}
\begin{equation}
\label{eq.emisdpeak}
\epsilon '_{\nu',\mathrm{p}} = 0.88 \cdot \frac{256}{27}\frac{p-1}{3p-1}\frac{\qe^3}{\me c^2} n'B',
\end{equation}
where $\qe$ is the elementary charge and $\me$ is the electron mass. $n'$ and $B'$ are the number density and the strength of the magnetic field in the shocked medium, respectively, which are given in the relativistic limit as follows \citep{BM}:
\begin{align}
\label{eq.nd}
n' &= 4\Gamma n_0, \\
\label{eq.ed}
e'_\mathrm{i} &= (\Gamma-1)n'\mpr c^2,\\
\label{eq.Bd}
B' &= \sqrt{8\pi \eB e'_\mathrm{i}} = \sqrt{32\pi \eB n_0 \Gamma(\Gamma -1)\mpr c^2}.
\end{align}
Here, $e'_\mathrm{i}$ is the internal energy density of the shocked fluid and $\eB$ is the energy conversion efficiency from shocked matter to magnetic field. We note that Equations~(\ref{eq.nd}) and (\ref{eq.ed}) approach the strong-shock limit of non-relativistic shock with the ratio of the specific heats $\hat{\gamma}=5/3$ in the limit of $\Gamma \rightarrow 1$. Hence we employ Equations~(\ref{eq.nd}), (\ref{eq.ed}), and (\ref{eq.Bd}) for both relativistic and non-relativistic regimes.
The two break frequencies are given by \citep{Granot99,Eerten10}
\begin{align}
\label{eq.numd}
\numd &=  \frac{3}{16}  \frac{\gamma _\mathrm{m}^{\prime 2}\qe B'}{\me c} = \frac{3}{16} \left[ \ee \frac{p-2}{p-1}\frac{\mpr}{\me}(\Gamma - 1) \right]^2 \frac{\qe B'}{\me c}, \\
\label{eq.nucd}
\nucd &=  \frac{3}{16}  \frac{\gamma _\mathrm{c}^{\prime 2}\qe B'}{\me c} = \frac{3}{16} \left[\frac{3\me c \Gamma}{4\sigma _\mathrm{T} \eB e_\mathrm{i}' t }\right]^2 \frac{\qe B'}{\me c},
\end{align}
where $\gamma _\mathrm{m}^\prime$ is the minimal Lorentz factor of the non-thermal electrons, $\gamma _\mathrm{c}^\prime$ is the characteristic Lorentz factor for cooling, $\ee$ is a model parameter that gives the energy conversion efficiency from shocked matter to the non-thermal electrons,
and $\sigma_\mathrm{T}$ is the cross section of Thomson scattering.

We emphasize that the local synchrotron emissivity depends on the shock energy $E$, since $\epsilon '_{\nu',\mathrm{p}}$, $\numd$, and $\nucd$ are functions of $E$ through $\Gamma$ given by Equation~(\ref{eq.BM-ST}). The rest frame frequency $\nu'$ also depends on $E$ via the Lorentz transformation of a given observed frequency $\nu$:
\begin{equation}
\label{eq.nud}
\nu' = \Gamma(1-\beta \mu)\nu .
\end{equation}

The observed flux density at an observer time $T$ and an observed frequency $\nu$ is given by integrating the emission coefficient of synchrotron radiation $j _\nu$ \citep{Granot99}:
\begin{equation}\label{Forigin}
F_{\nu}(T)=\frac{1}{D^2}\int_0^{\thj} \diff \theta \sin \theta \left. \int _0^{2\pi} \diff \phi \int _0^\infty \diff r r^2 j _\nu \right|_{t = T+\mu r/c}, 
\end{equation}
where $r$ is the radius from the centre of the spherically expanding ejecta, $\theta$ is the angle measured from the jet symmetric axis, and $\phi$ is the azimuthal angle measured from the observer direction (i.e., $\phi = 0$ for the observer). $D$ is the luminosity distance to the source and the effect of cosmological redshift $z$ is neglected for simplicity ($z\sim 0$). $\thj$ is the jet half-opening angle. We choose $T=0$ as the arrival time of a photon emitted at the origin at $t=0$. 
Then, 
\begin{equation}
\label{eq.t_generic}
t = T+\frac{\mu r}{c}
\end{equation}
is the laboratory time at each position when the emitted photons reach the observer at $T$, where
\begin{equation}
\label{eq.mu}
\mu = \sin \theta \sin \view \cos \phi + \cos \theta \cos \view
\end{equation}
is the cosine of the angle spanned by the radial vector and the line of sight with $\view$ being the viewing angle measured from the jet axis. We note that $j_\nu(\theta, \phi)$ generally has a directional dependence in the laboratory frame.

Equation~(\ref{Forigin}) can be further reduced to a simpler form: The emission coefficient can be written as $j _\nu = j'_{\nu'}/[\Gamma^2(1-\beta \mu)^2] = \emisd/[4\pi \Gamma^2(1-\beta \mu)^2]$ by assuming that the synchrotron radiation is isotropic in the fluid rest frame. Furthermore, the thin-shell and relativistic shock approximations reduce the integration with respect to $r$ as follows \citep{Eerten10}:
\begin{align}
F_{\nu}(T)&\sim\frac{1}{4\pi D^2}\int_0^{\thj} \diff \theta \sin \theta \int _0^{2\pi} \diff \phi 
\left. \frac{R^2\Delta R\emisd}{\Gamma^2(1-\beta \mu)^2} \right|_{t = T+\mu R/c},\\
\label{eq.Fbasic}
&\sim \frac{1}{4\pi D^2}\int_0^{\thj} \diff \theta \sin \theta \int _0^{2\pi} \diff \phi 
\nonumber \\ 
& \qquad \qquad \qquad \times \left. \frac{R^3\emisd}{12\Gamma^4(1 - \beta_\sh \mu)(1 - \beta \mu)^2} \right| _{t = T+\mu R/c},
\end{align}
where $\Delta R \sim R/[12\Gamma^2(1 -\beta_\sh \mu)]$ is the width of the shocked region that emits the photons observed at $T$. We note that the integrand depends on $\phi$ through $\mu$ and $t$ for an off-axis observer ($\view \ne 0$). The laboratory time $t$ corresponding to a given observer time $T$ is found for each position $(\theta, \phi)$ by solving
\begin{equation}
\label{eq.t}
t = T + \frac{\mu R}{c}
\end{equation}
with Equation~(\ref{eq.R}), where we substituted $r=R$ in Equation~(\ref{eq.t_generic}). The laboratory time that satisfies Equation~(\ref{eq.t}) can be numerically found with a standard root-finding algorithm.

Afterglow light curves for an observed frequency $\nu$ are synthesized by Equation~(\ref{eq.Fbasic}) with Equations~(\ref{eq.BM-STsh})-(\ref{eq.nud}), (\ref{eq.mu}), and (\ref{eq.t}) for a given parameter set of $\{E(\theta), n_0, \eB, \ee, p, \thj, \view, D \}$, where $E(\theta)$ is the angle dependence of the isotropic equivalent energy of a given structured jet. 

\begin{figure}
	\includegraphics[bb = 0 0 461 346, width = \columnwidth]{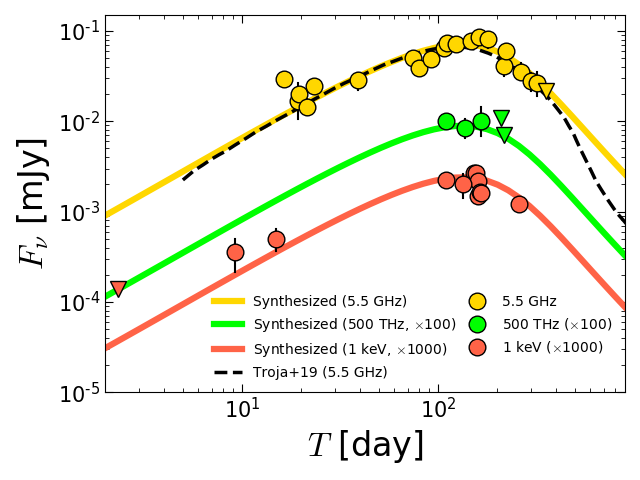}
	\caption{Afterglow light curves for radio (yellow line), optical (green line), and X-ray (red line) synthesized by Equation~(\ref{eq.Fbasic}) with a Gaussian jet given by Equation~(\ref{eq.jet_Gaussian}) with the model parameters shown below the equation. Also shown are the observed afterglow of GRB170817A (points) and the best-fitting radio light curve taken from \citet{Troja19} (black dashed line). The lower triangles are upper limits. The data points for radio were taken from Figure~4 in \citet{Troja19}, which used the data in \citet{Hallinan17,Lyman18,Troja18a,Margutti18,Mooley18,Alexander18,Piro19}. The data points for optical and X-ray were collected from \citet{Lyman18,Margutti18,DAvanzo18,Alexander18,Piro19}.}
	\label{fig.demonstration}
\end{figure}

We demonstrate an afterglow light curve produced by a Gaussian jet as an example, whose energy distribution is given by
\begin{equation}
\label{eq.jet_Gaussian}
E(\theta) = \Ec \exp\left(-\frac{\theta^2}{2\thc^2}\right),
\end{equation}
where $\Ec$ is the isotropic equivalent energy measured at the jet axis and $\thc$ is the standard deviation of the Gaussian.
Figure~\ref{fig.demonstration} shows the synthesized light curves for $\log(\Ec/\mathrm{erg})=52.8$, $\thc=0.059$, $\log (n_0/\mathrm{cm}^{-3}) = -2.28$, $\log \eB = -4.68$, $\log \ee = -1.39$, $p=2.17$, $\thj = 0.61$, $\view=0.387$, and $D=41$~Mpc, which is the distance to the host galaxy of GRB~170817A \citep{D,Cantiello18}. We also depict the observed fluxes of the afterglow of GRB~170817A and the fitted light curve taken from \citet{Troja19} for comparison, who also assumed the same Gaussian jet to synthesize the light curve.\footnote{The parameter values in \citet{Troja19} are $\log (n_0/\mathrm{cm}^{-3}) = -2.51$, $\log \eB = -4$, $\log \ee = -1.39$, $p=2.1681$, $\thj = 0.61$, and $\view=0.38$, with unclarified distance $D$. We modified these values to fit the light curve, since the afterglow becomes overluminous by a factor of $\sim2.2$--$2.6$ in the rising portion $6\le T/\mathrm{day}\lesssim130$ for these original values.} As shown in Figure~\ref{fig.demonstration}, the rising part of the synthesized radio light curve is consistent with that of \citet{Troja19}. The deviation of the light curves in late time would be due to our ignorance of the sideway expansion of the jet, which was taken into account in \citet{Troja19} and becomes important as the jet is decelerated to non-relativistic speeds \citep{KG03}. However, this effect is not important for our purpose, since our inversion formula uses only a rising part of light curves before the jet break as explained later.

\subsection{Derivation of the inversion formula} \label{sec.Inv}
Based on the above basic equations, we derive the inversion formula to inversely reconstruct the jet energy distribution $E(\theta)$ from a given afterglow light curve of an off-axis GRB. In the inversion process, we do not assume the functional form of $E(\theta)$ while we fix the values of the other parameters, $n_0$, $\eB$, $\ee$, $p$, $\thj$, $\view$, and $D$ (See Section~\ref{sec.Inv-edge} for fixing these parameters). This is an inverse problem to solve the integral equation, Equation~(\ref{eq.Fbasic}), for $E(\theta)$. However, it is a non-trivial task, since the integrand depends on $E(\theta)$ in a complicated form and it cannot be split into a kernel that does not depend on $E(\theta)$ and the other part that depends on $E(\theta)$. Hence, we propose a novel method to solve the integral equation for $E(\theta)$ by properly approximating Equation~(\ref{eq.Fbasic}) from a physical point of view.

\subsubsection{Essence of the method} \label{sec.Inv-essence}
Our idea that easily solves Equation~(\ref{eq.Fbasic}) for $E(\theta)$ is to approximate Equation~(\ref{eq.Fbasic}) as follows:
\begin{align}
\label{eq.F}
F_\nu (T) &\sim \frac{1}{4\pi D^2}\int_{\Theta(T)}^{\thj} \diff \theta \sin \theta \int _0^{2\pi} \diff \phi \nonumber \\
& \qquad \qquad \qquad \times \left. \frac{R^3\emisd}{12\Gamma^4(1 - \beta_\sh \mu)(1 - \beta \mu)^2} \right| _{t = T+\mu R/c},
\end{align}
where we reduced the interval of integration with respect to $\theta$ from $[0, \thj]$ to $[\Theta(T), \thj]$ by introducing a cutoff angle $\Theta(T)$, where $\Theta(T)$ monotonically decreases with $T$. It is the essential point in our method to introduce $\Theta(T)$ here, while the justification of the approximation is given in the following paragraphs and a specific functional form of $\Theta(T)$ is given in the next subsection. To explain the idea, let us consider the observed flux at an observer time $T + \delta T$, which is a slightly proceeded time for an arbitrary time $T$. The observed flux $F_\nu(T+\delta T)$ given by Equation~(\ref{eq.F}) is the sum of the two different contributions: One is the new contribution from the inner region that becomes observable at $T+\delta T$, $\Theta(T+\delta T) \le \theta < \Theta(T)$, and the other is the contribution from the outer region that has been observable so far, $\Theta(T) \le \theta \le \thj$. The latter part can be calculated, if one already knows the energy distribution for $\Theta(T) \le \theta \le \thj$ and the other model parameters $\{n_0, \eB, \ee, p, \thj, \view, D \}$. Then, in principle, the energy contained in the newly observable region, $\Theta(T+\delta T) \le \theta < \Theta(T)$, can be estimated from the rest of the observed flux, $F_\nu(T+\delta T)-\mathrm{(the\ latter\ contribution)}$. By iterating this procedure for a given time interval, we can obtain the jet structure from a given light curve. In Sections~\ref{sec.Inv-inv} and \ref{sec.Inv-edge}, we specify $\Theta(T)$ and give the detail of the inversion procedure.

\begin{figure*}
	\begin{tabular}{cc}
		\begin{minipage}{0.45\hsize}
			\begin{center}
				\includegraphics[bb = 0 0 384 338, width=\textwidth]{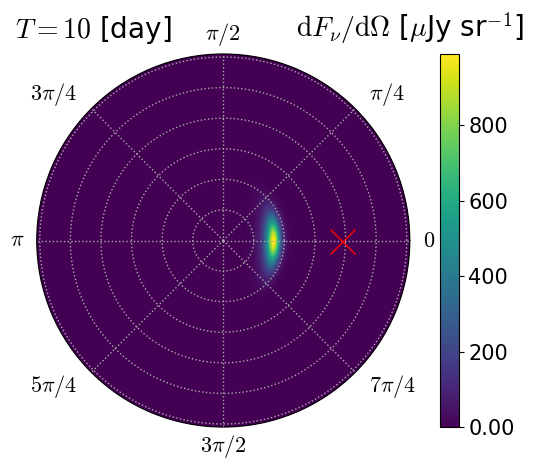}
			\end{center}
		\end{minipage} &
		\begin{minipage}{0.45\hsize}
			\begin{center}
				\includegraphics[bb = 0 0 384 338, width=\textwidth]{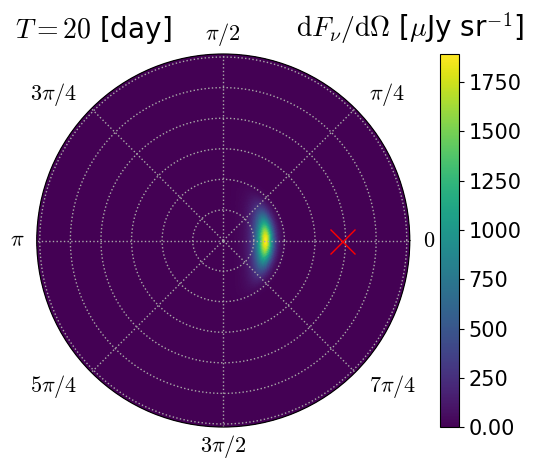}
			\end{center}
		\end{minipage} \\
		\begin{minipage}{0.45\hsize}
			\begin{center}
				\includegraphics[bb = 0 0 384 338, width=\textwidth]{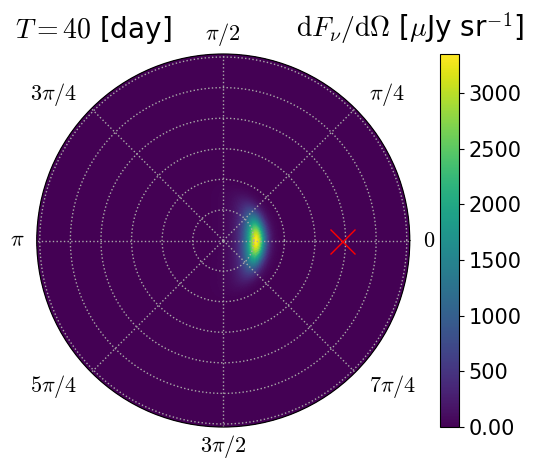}
			\end{center}
		\end{minipage} &
		\begin{minipage}{0.45\hsize}
			\begin{center}
				\includegraphics[bb = 0 0 384 338, width=\textwidth]{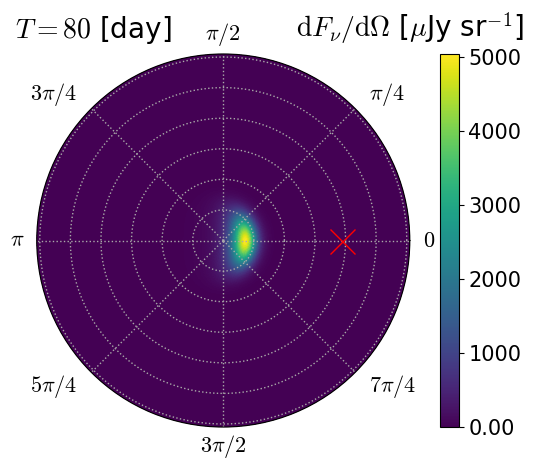}
			\end{center}
		\end{minipage} \\
		\begin{minipage}{0.45\hsize}
			\begin{center}
				\includegraphics[bb = 0 0 384 338, width=\textwidth]{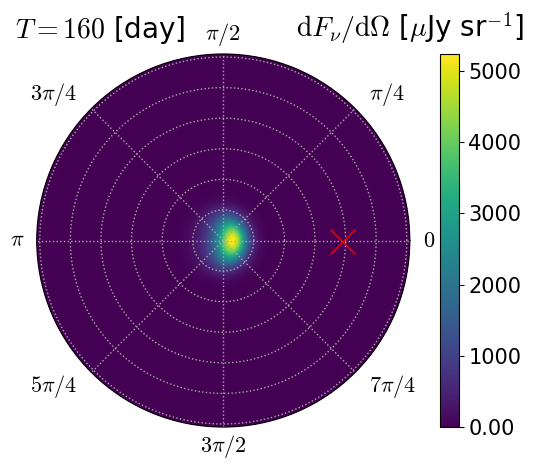}
			\end{center}
		\end{minipage} &
		\begin{minipage}{0.45\hsize}
			\begin{center}
				\includegraphics[bb = 0 0 384 338, width=\textwidth]{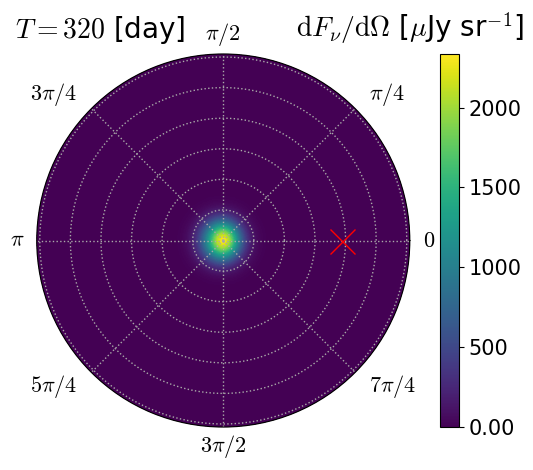}
			\end{center}
		\end{minipage} 
	\end{tabular}
	\caption{Colour maps of the contributing flux per unit solid angle $\diff F_\nu/\diff \Omega$ ($\nu=5.5$~GHz) on $(\theta, \phi)$-polar coordinates for the Gaussian jet given by Equation~(\ref{eq.jet_Gaussian}) with the parameters shown below the equation. The dotted-circle grids indicate $\theta = 0.1,\ 0.2,\ 0.3,\ 0.4,\ 0.5,\ 0.6$ from the innermost line to the outermost one, respectively, while dotted-radial grids indicate the $\phi$ coordinate designated outside. The outside edge corresponds to the jet truncation angle $\thj = 0.61$. The observer direction $(\view, 0)=(0.387,0)$ is marked as a red cross. The observer time $T$ is displayed in the top left corner in each panel.}
	\label{fig.dist_Lv0}
\end{figure*}

We now justify Equation~(\ref{eq.F}). Most importantly, we point out that the observed flux at each time is contributed only from a limited region of the jet, at least in early phase, mainly because only a fraction of the emitted photons reaches the off-axis observer due to relativistic beaming effects. We illustrate this idea by taking the Gaussian jet used in Section~\ref{sec.review} as an example: Figure~\ref{fig.dist_Lv0} shows the evolution of the surface brightness,
where each panel displays the colour map of the contributing flux per unit solid angle for a given observer time $T$ and $\nu=5.5$~GHz:
\begin{equation}
\frac{\diff F_\nu}{\diff \Omega}= \frac{1}{4\pi D^2} \left. \frac{R^3\emisd}{12\Gamma^4(1 - \beta_\sh \mu)(1 - \beta \mu)^2} \right| _{t = T+\mu R/c}, 
\end{equation}
which is obtained by differentiating Equation~(\ref{eq.Fbasic}) with respect to the solid angle $\Omega$. As seen in each panel, only a limited region contributes to the observed flux and the luminous region gradually moves toward the jet axis as time passes. The side near to the off-axis observer is more luminous, since the emission from the other side is de-beamed by relativistic beaming effects. In fact, there is a strong correlation between the luminous region with large $\diff F_\nu/\diff \Omega$ and the so-called beaming factor:
\begin{equation}
\delta :=\frac{1}{\Gamma(1-\beta \mu)},
\end{equation}
as shown in Equation~(\ref{eq.Fpropto}) below. Hence, the inner region is not visible for an off-axis observer in early phase, since the emission is strongly de-beamed from the observer. The inner region gradually becomes visible as the shock is decelerated, which shifts the luminous region in Figure~\ref{fig.dist_Lv0} inward with time. The above consideration safely reduces the interval of integration with respect to $\theta$ in Equation~(\ref{eq.Fbasic}) from $[0, \thj]$ to $[\Theta(T), \thj]$ as given in Equation~(\ref{eq.F}), where $\Theta(T)$ corresponds to the inner edge of the luminous region.

\begin{figure*}
	\begin{tabular}{cc}
		\begin{minipage}{0.45\hsize}
			\begin{center}
				\includegraphics[bb = 0 0 384 341, width=\textwidth]{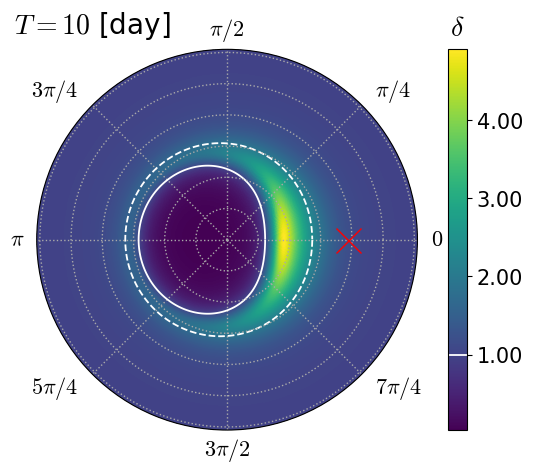}
			\end{center}
		\end{minipage} &
		\begin{minipage}{0.45\hsize}
			\begin{center}
				\includegraphics[bb = 0 0 384 341, width=\textwidth]{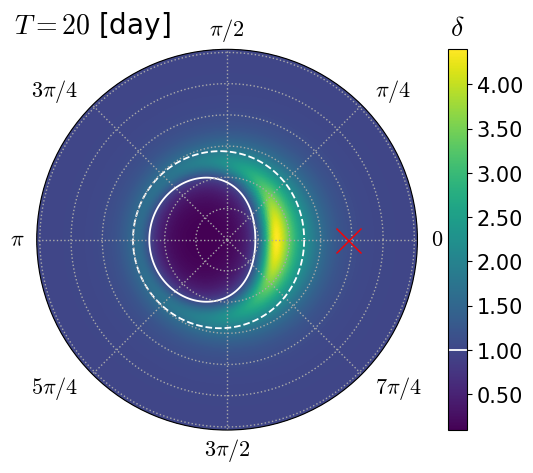}
			\end{center}
		\end{minipage} \\
		\begin{minipage}{0.45\hsize}
			\begin{center}
				\includegraphics[bb = 0 0 384 341, width=\textwidth]{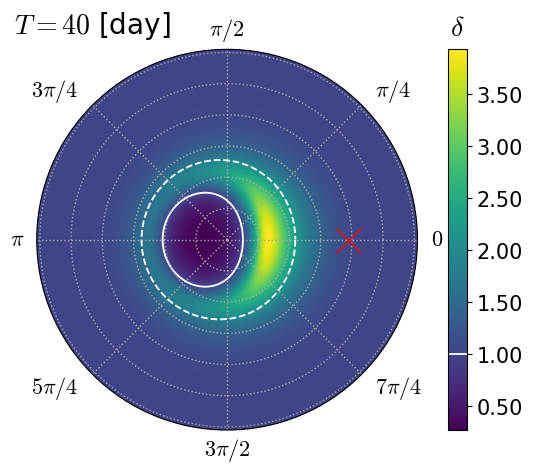}
			\end{center}
		\end{minipage} &
		\begin{minipage}{0.45\hsize}
			\begin{center}
				\includegraphics[bb = 0 0 384 341, width=\textwidth]{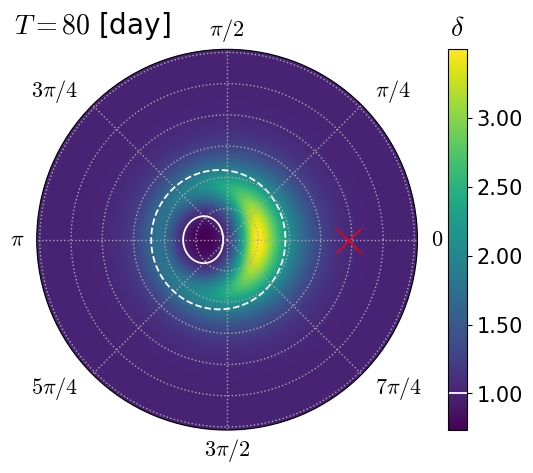}
			\end{center}
		\end{minipage} \\
		\begin{minipage}{0.45\hsize}
			\begin{center}
				\includegraphics[bb = 0 0 384 341, width=\textwidth]{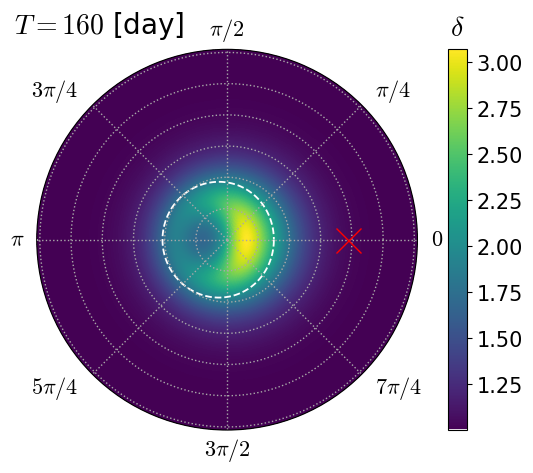}
			\end{center}
		\end{minipage} &
		\begin{minipage}{0.45\hsize}
			\begin{center}
				\includegraphics[bb = 0 0 384 341, width=\textwidth]{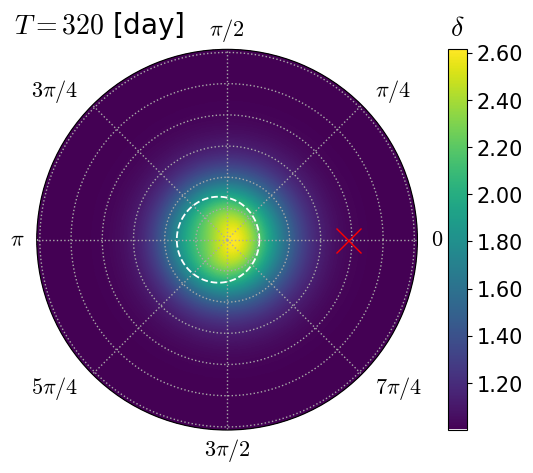}
			\end{center}
		\end{minipage} 
	\end{tabular}
	\caption{Same as Figure~\ref{fig.dist_Lv0} but for the beaming factor $\delta:=1/[\Gamma(1-\beta \mu)]$. The thick solid and dashed lines indicate $\delta=1$ and $\Gamma \beta = 1$, respectively.}
	\label{fig.gbf_Lv0}
\end{figure*}

We can show the following proportionality for fixed $T$ and $\nu$ in the relativistic limit: 
\begin{equation}
\label{eq.Fpropto}
\frac{\diff F_\nu}{\diff \Omega} \propto E^{(3p+5)/10}\delta^{2(10-p)/5},
\end{equation}
where we employed $R\sim ct$, $t \sim T/(1 - \beta \mu)$, $\Gamma \propto E^{1/2}t^{-3/2}$, and $1-\beta_\sh \mu \sim 1 - \beta \mu$ for $\Gamma \gg 1$ and $\beta \sim 1$. We also used the relativistic limit for the slow cooling, $\emisd \propto \Gamma ^{1+p}(1 -\beta \mu)^{-(p-1)/2}$ ($\numd < \nu' < \nucd$), which is relevant in this case. Figure~\ref{fig.gbf_Lv0} manifests this correspondence, which shows the distribution of the beaming factor $\delta$ (cf.~Figure~\ref{fig.dist_Lv0}). The thick dashed line shows the region where $\Gamma \beta = 1$, which is a diagnostic boundary between relativistic ($\Gamma \beta > 1$) and non-relativistic ($\Gamma \beta < 1$) regions. The jet edge region with $\theta \gtrsim 0.35$ becomes non-relativistic earlier than $T=10$~days, since the jet energy steeply decays toward the edge owing to the given
Gaussian structure, and hence does not much contribute to the observed light curve for $T\ge 10$~days even with $\delta > 1$. The relativistic region is divided to the regions where the emission is beamed to/away from the observer direction ($\delta > 1$ and $\delta < 1$, respectively). The strong dependence of $\diff F_\nu/\diff \Omega \propto \delta^{2(10-p)/5}$ on $\delta$ enhances the contrast as shown in Figure~\ref{fig.dist_Lv0}, whereas the peak position could be slightly shifted inward or outward because of the distribution of $E(\theta)$. As time passes and the shock is decelerated, the relativistic region shrinks and the de-beamed region with $\delta < 1$ disappears as shown in the bottom panels in Figure~\ref{fig.gbf_Lv0}. 

\subsubsection{Inversion equation} \label{sec.Inv-inv}
In this paper, $\Theta(T)$ is given by $\theta$ that satisfies the following equation:
\begin{equation}
\label{eq.cond-th}
\theta + \left. \frac{\fb}{\Gamma(t,\theta)}\right|_{t = T+\mu R/c,\ \phi=0} = \view,
\end{equation}
where we introduced a factor $\fb$ to expand the size of the beaming cone, which is usually given by $1/\Gamma$. Note that a fraction of $\fb^2/(1+\fb^2)$ of the photons emitted isotropically in the rest frame is beamed to the cone with a half-opening angle $\thb := \fb/\Gamma$.\footnote{The condition that the observer direction is on the edge of the cone with a half-opening angle $\thb$ is given by $\mu = \cos \thb$ or, equivalently, $\cos \phi  = (\cos \thb - \cos \theta \cos \view)/(\sin \theta \sin \view)$, which is reduced to Equation~(\ref{eq.cond-th}) for $\phi=0$.} 
Thus, Equation~(\ref{eq.cond-th}) gives the polar angle inside which a fraction more than $\fb^2/(1+\fb^2)$ of the emitted synchrotron photons do not reach the observer due to relativistic beaming effects. In the case of $\fb=7$, for example, more than 98 per cent of the emitted photons does not reach the observer for $\theta < \Theta(T)$, which is neglected in the integration in Equation~(\ref{eq.F}). Note that larger $\fb$ gives smaller $\Theta(T)$ for fixed jet structure and $T$ (See Appendix~\ref{app.fb}).

The inner truncation angle $\Theta(T)$ can be given in a more specific form by Equation~(\ref{eq.cond-th}).
Since $\theta = \Theta(T)$ lies in the relativistic region during the reconstruction, it is a good approximation to use the Blandford-McKee solution in Equation~(\ref{eq.cond-th}):
\begin{align}
\label{eq.Gamma_sh-BM}
\Gamma_\sh (t, \theta) &= C_\mathrm{BM}(\theta) t^{-3/2},\\
\label{eq.Gamma-BM}
\Gamma (t, \theta) &= \frac{1}{\sqrt{2}}C_\mathrm{BM}(\theta) t^{-3/2},
\end{align}
which are also obtained by formally putting $\beta_\sh = \beta = 1$ and neglecting the Sedov-Taylor terms in Equations~(\ref{eq.BM-STsh}) and (\ref{eq.BM-ST}).
Note here that $C_\mathrm{BM}(\theta)$ depends on $\theta$ through $E(\theta)$ [See Equation~(\ref{eq.C_BM})]. Here, the laboratory time $t$ for $\theta = \Theta(T)$ is given by Equations~(\ref{eq.mu}), (\ref{eq.t}), and (\ref{eq.cond-th}) as follows:
\begin{align}
t &= T + \frac{R}{c}\cos \left(\frac{\fb}{\Gamma} \right),\\
&\sim T + t\left(1 - \frac{1}{16\Gamma^2}\right)\left[1 - \frac{1}{2} \left(\frac{\fb}{\Gamma} \right)^2\right],\\
&\sim T+t\left[1 - \frac{1}{16\Gamma^2} -\frac{1}{2} \left(\frac{\fb}{\Gamma} \right)^2 \right],\\
\label{eq.t-bound}
\Rightarrow t&\sim\frac{16\Gamma^2T}{1+8\fb^2},
\end{align}
where we employed $R \sim ct [1 - 1/(16\Gamma^2)]$ in the second line, which follows from Equations~(\ref{eq.R}), (\ref{eq.Gamma_sh-BM}), and (\ref{eq.Gamma-BM}).
Combining Equations~(\ref{eq.C_BM}), (\ref{eq.cond-th}), (\ref{eq.Gamma-BM}), and (\ref{eq.t-bound}), we obtain
\begin{equation}
\label{eq.Theta}
\Theta = \view - A E^{-1/8}(\Theta) T^{3/8},
\end{equation}
where $A$ is a constant given by
\begin{equation}
\label{eq.A}
A:= 4\fb \left[\frac{\pi n_0 \mpr c^5}{17(1 + 8\fb^2)^3}\right]^{1/8}.
\end{equation}
The changing rate of $\Theta(T)$ is obtained by differentiating Equation~(\ref{eq.Theta}) with respect to $T$:
\begin{equation}
\label{eq.dThetadT}
\frac{\diff \Theta}{\diff T} = -\frac{3(\view - \Theta)}{8T}\left(1 - \frac{\view - \Theta}{8}\frac{\diff \ln E}{\diff \Theta}\right)^{-1}.
\end{equation}
Thus, $\Theta(T)$ monotonically decreases with time as long as $\diff E/\diff \theta < 8E/(\view - \theta)$. 
The time when the jet axis becomes visible, $T_\mathrm{f}$, is given by $\Theta(T_\mathrm{f}) = 0$ as follows:
\begin{equation}
\label{eq.Tf}
T_\mathrm{f} = 60.2\ \mathrm{day} \left(\frac{\Ec}{10^{53}\ \mathrm{erg}}\right)^{1/3}\left(\frac{n_0}{10^{-2}\ \mathrm{cm}^{-3}}\right)^{-1/3}\left(\frac{\view}{0.4} \right)^{8/3} , 
\end{equation}
where $\Ec = E(0)$ is the isotropic equivalent energy measured at the jet axis and we employed $\fb = 7$. It would be interesting to note here that $T_\mathrm{f}$ is not the same as the observer time for the peak of a light curve, $T_\mathrm{p}$. Assuming that the peak time corresponds to the time for jet break, one obtains $T_\mathrm{p} \propto  E_\mathrm{tot}^{1/3}n_0^{-1/3}\view ^2$ \citep{Nakar02,GNP19}, where $E_\mathrm{tot}$ is the total jet energy and the dependence is valid for an off-axis observer whose viewing angle is much larger than the jet core angle size. As shown later in Section~\ref{sec.results}, $T_\mathrm{f}$ is indeed smaller than $T_\mathrm{p}$ in the considered cases.

Since $T_\mathrm{f}$ should be larger than a given initial time $T_0$, we obtain the off-axis condition on the viewing angle $\view$ from Equation~(\ref{eq.Tf}):
\begin{equation}
	\label{eq.off-axis}
	\view \ge \theta_\mathrm{v,min} = 0.20 \left(\frac{\Ec}{10^{53}\ \mathrm{erg}}\right)^{-1/8}\left(\frac{n_0}{10^{-2}\ \mathrm{cm}^{-3}}\right)^{1/8}\left(\frac{T_0}{10~\mathrm{day}} \right)^{3/8}.
\end{equation}
Our inversion method cannot be applied for $\view < \theta_\mathrm{v,min}$, for which the entire region of the jet has been visible to the observer from the initial time.

Finally, we obtain the following inversion formula for reconstructing the jet energy distribution $E(\theta)$ by differentiating Equation~(\ref{eq.F}) with respect to $T$ and employing Equation~(\ref{eq.dThetadT}):
\begin{align}
\label{eq.inversion}
\frac{\diff \ln E}{\diff \Theta} &= \frac{8}{\view - \Theta} -\frac{3 K(T,\Theta,E(\Theta))}{F_\nu(T)} \left[\frac{\diff \log F_{\nu}}{\diff \log T}(T) \right. \nonumber \\
&\qquad \qquad \qquad \quad \left. - \frac{T}{F_\nu(T)}\int _{\Theta}^{\thj} \diff \theta \frac{\diff K}{\diff T}(T,\theta,E(\theta))\right]^{-1},
\end{align}
where $T$ is related to $\Theta$ by Equation~(\ref{eq.Theta}). $K$ is defined by
\begin{align}
\label{eq.Kbefore}
& K(T, \theta, E(\theta)) \nonumber \\ &:= \frac{1}{4\pi D^2} \int _0^{2\pi} \diff \phi \left. \frac{\sin \theta R^3\emisd}{12\Gamma^4(1 - \beta_\sh \mu)(1 - \beta \mu)^2} \right| _{t = T+\mu R/c},\\
\label{eq.K}
&\sim \frac{1}{4\pi D^2} \int _0^{2\pi} \diff \phi \left. \frac{\sin \theta \Rs^3\emisd}{12\Gamma^4(1 - \beta_\sh \mu)(1 - \beta \mu)^2} \right| _{t = \ts(T, \theta, \phi; E(\theta))}, 
\end{align}
where $\ts$ in the second line is an approximated solution for Equation~(\ref{eq.t}), which is explicitly given by $T$, $\theta$, $\phi$, and $E(\theta)$ as in Equation~(\ref{eq.ts}). This approximation with relativistic limits is not necessary for inversion but saves the computational time to numerically solve Equation~(\ref{eq.t}). We also replace $R$ to $\Rs$ given by Equation~(\ref{eq.Rs}), which is a reasonable approximation for the shock radius in relativistic regions. The Lorentz factors and velocities of the shock and shocked fluid in Equation~(\ref{eq.K}) are given by Equations~(\ref{eq.Gammabetash}) and (\ref{eq.Gammabeta}), which also follow from the relativistic limit. Throughout the paper, we always use Equation~(\ref{eq.K}) for $K$ instead of Equation~(\ref{eq.Kbefore}). Note that the approximated observed flux is written by using $K$ as
\begin{equation}
\label{eq.Fapp}
F_\nu(T) = \int_{\Theta(T)}^{\thj} \diff \theta K(T,\theta,E(\theta)).
\end{equation}
Furthermore, we use the following synchrotron emissivity in the inversion process instead of Equation~(\ref{eq.slowcooling}) for simplicity:
\begin{equation}
\label{eq.emisd}
\emisd = \epsilon'_{\nu',\mathrm{p}} \left( \frac{\nu '}{\numd}\right)^{-(p-1)/2},
\end{equation} 
which is sufficient to explain the observed afterglow spectrum of GRB~170817A. Then, $\diff K/\diff T$ can be calculated by using the chain rule:
\begin{equation}
\label{eq.dKdT}
\frac{\diff K}{\diff T} = \frac{\diff K}{\diff \ts}\frac{\diff \ts}{\diff T},
\end{equation}
where the explicit forms of $\diff K/\diff \ts$ and $\diff \ts/\diff T$ are given by Equations~(\ref{eq.dKdts}) and (\ref{eq.dtsdT}), respectively. 

It is important to emphasize here that the right-hand side of Equation~(\ref{eq.inversion}) depends only on the energy distribution $E(\theta)$ for $\Theta(T) \le \theta \le \thj$ and is independent of $E(\theta)$ for $\theta < \Theta(T)$. Hence, once the jet energy distribution $E(\theta)$ is given for $\Theta_0 \le \theta \le \thj$, the jet structure is uniquely reconstructed by integrating Equation~(\ref{eq.inversion}) from $\Theta_0$ to $\Theta=0$ inward for given light curve and parameter set $\{n_0, \eB, \ee, p, \thj, \view, D, \fb \}$, where $\Theta_0 := \Theta(T_0)$ is the cutoff angle for a given initial time $T_0$. 
The way to give $E(\theta)$ ($\Theta_0 \le \theta \le \thj$) and $\{n_0, \eB, \ee, p, \thj, \view, D, \fb \}$ is explained in the next subsection.

We note that we can choose either of $\Theta$ or $T$ as an independent variable, while the other is then determined by Equation~(\ref{eq.Theta}), when we numerically integrate the differential Equation~(\ref{eq.inversion}). In the remainder of the paper, we choose $\Theta$ as an independent variable except for $\Theta_0$, which is fixed by a given $T_0$ through Equation~(\ref{eq.Theta}) as explained in the next subsection. We divide the interval $[0, \Theta_0]$ by a mesh with equally-spaced $N$ grid points : $0 = \Theta_{N-1} < \Theta_{N-2} < \cdots < \Theta_1 < \Theta_0$ and employ the 4-th order Runge-Kutta method to integrate Equation~(\ref{eq.inversion}). The corresponding observer times, $T_\mathrm{f} = T_{N-1} > T_{N-2} > \cdots > T_1 > T_0$, are not {\it a priori} known except for $T_0$, since they are determined by Equation~(\ref{eq.Theta}) and, hence, depends on the energy distribution $E(\theta)$ that is to be obtained in the inversion process. 

\subsubsection{Constraints on model parameters} \label{sec.Inv-edge}
We should specify the model parameters $\{n_0, \eB, \ee, p, \thj, \view, D, \fb \}$ and the jet energy distribution $E(\theta)$ in the jet edge part $\Theta_0 \le \theta \le \thj$ to integrate Equation~(\ref{eq.inversion}).
These are not determined by the inversion process but should be obtained in some way beforehand. We adopt $\fb = 7$ in this paper, which turns out to be a reasonable value as shown in Section~\ref{sec.results_demonstration}. The viewing angle $\view$ is constrained by superluminal apparent motions of afterglow images and/or gravitational wave signals. The spectral index $p$ is obtained by multi-frequency observations. The luminosity distance is obtained by the host galaxy and/or gravitational wave signals. The other parameters $n_0$, $\eB$, and $\ee$ can be determined if the absorption, characteristic, and cooling break frequencies are obtained. Otherwise, it is generally difficult to determine these parameters because they are degenerate. It would be worth noting that $\ee$ is typically $\sim 0.1$ in observations \citep[e.g.,][]{KZ15} and simulations \citep[e.g.,][]{SS11}.

We can further give constraints on these parameters by using the light curve at the initial observer time $T_0$. These parameters should satisfy the following condition by definition:
\begin{equation}
\label{eq.constraint1}
F_\nu(T_0) = \int _{\Theta_0}^{\thj} \diff \theta K(T_0, \theta, E(\theta)).
\end{equation}
In addition, if the light curve is smooth at $T_0$, 
the parameters are also constrained by the following equations for $k =1,2,3,\cdots$: 
\begin{equation}
\label{eq.constraint_k}
\frac{\diff^kF_\nu}{\diff T^k}(T_0) =\left.  \frac{\diff^k}{\diff T^k} \int _{\Theta(T)}^{\thj} \diff \theta K(T, \theta, E(\theta)) \right|_{T=T_0}.
\end{equation}
In this paper, we assume that the first derivative of $F_\nu$ always exits at $T_0$ and the parameters satisfy Equation~(\ref{eq.constraint_k}) for $k=1$:
\begin{equation}
\label{eq.constraint2}
\frac{\diff F_\nu}{\diff T}(T_0) = \frac{\diff \Theta}{\diff T}(T_0) K(T_0, \Theta_0, E(\Theta_0))
+ \int _{\Theta_0}^{\thj} \diff \theta \frac{\diff K}{\diff T}(T_0, \theta, E(\theta)) .
\end{equation}
We can then reduce two degrees of freedom
(in particular for $E(\theta)$ at the jet edge part)
by Equations~(\ref{eq.constraint1}) and (\ref{eq.constraint2}).

Practically, as applied in Section~\ref{sec.results}, we first fix the parameters $\{n_0, \eB, \ee, \thj, p, \view, D, \fb \}$ to some values and put the energy distribution in the jet edge part as a function with two free parameters, $a$ and $b$: $E=E(\theta, a, b)$ for $\Theta_0 \le \theta \le \thj$. Equations~(\ref{eq.constraint1}) and (\ref{eq.constraint2}) then give $a$ and $b$ for a given light curve at $T_0$. The values of $a$ and $b$ satisfying Equations~(\ref{eq.constraint1}) and (\ref{eq.constraint2}) are numerically found with iteration by a root-finding algorithm. We note here that $\Theta _0$ is a function of $a$ and $b$, since $\Theta_0$ depends on $E(\Theta_0, a, b)$ as given by Equation~(\ref{eq.Theta}). Hence, $\Theta_0$ in Equations~(\ref{eq.constraint1}) and (\ref{eq.constraint2}) changes with $a$ and $b$ in the iteration process.

We also note that model parameters are constrained after inversion as follows. The inversion formula use only a portion of a given light curve from $T=T_0$ to $T=T_\mathrm{f}$ in Equation~(\ref{eq.Tf}). Thus, we forwardly synthesize a light curve by using the reconstructed jet structure to check the consistency in the other time domain $T>T_\mathrm{f}$. If the synthesized light curve does not match the given light curve, we should change parameters and run the inversion process again. In this paper, we adjust $n_0$ and $\eB$ to make the peak time and peak flux consistent with the observed ones.

According to \citet{Nakar02,GNP19}, the peak time $\Tp$ and the peak flux $\Fp$ roughly obey the following scaling laws:
\begin{align}
\label{eq.Tp}
\Tp &\propto E_\mathrm{tot}^{1/3} n_0^{-1/3},\\
\label{eq.Fp}
\Fp &\propto E_\mathrm{tot} n_0^{(p+1)/4} \eB^{(p+1)/4} \ee^{p-1},
\end{align}
for fixed $p$, $\thj$, $\view$, $\nu$, and $D$.\footnote{Note that the meaning of the symbol $\thj$ in \citet{GNP19} is not the jet's truncation angle but the angle in which most of the jet's energy is contained.}
There are only two constraints, Equations~(\ref{eq.Tp}) and (\ref{eq.Fp}), for four unknowns, $E_\mathrm{tot}$, $n_0$, $\eB$, and $\ee$. Hence, we cannot fully determine the parameter values.

\subsubsection{Summary of the inversion method}
Starting from the standard theory of GRB afterglows and self-similar solutions of relativistic blast waves, 
we obtained the inversion formula, Equation~(\ref{eq.inversion}). The jet energy distribution $E(\theta)$ is inversely reconstructed by integrating Equation~(\ref{eq.inversion}) from $\Theta = \Theta_0$ to the jet axis ($\Theta = 0$). The inversion procedure is given as follows for a given light curve of a GRB afterglow $F_\nu(T)$ $(T\ge T_0)$, where $\nu$ is an observed frequency and $T_0$ is a given initial time in the observer frame.

\begin{enumerate}
	\item
	We specify the parameter values of $\{n_0, \eB, \ee, p, \thj, \view, D, \fb \}$, where $\fb$ is the parameter that defines the observable region by Equation~(\ref{eq.cond-th}).
	\item
	We assume the jet energy distribution in the jet edge part, $E(\theta, a, b)$ $(\Theta_0 \le \theta \le \thj)$, where $a$ and $b$ are free parameters in a given function and $\Theta_0(T_0, a, b)$ is the innermost angle of the observable region at $T_0$ given by Equations~(\ref{eq.Theta}) and (\ref{eq.A}). The free parameters $a$ and $b$ are determined so as to satisfy Equations~(\ref{eq.constraint1}) and (\ref{eq.constraint2}), which are constraints given by the observed flux $F_\nu(T_0)$ and its slope $\diff F_\nu/\diff T(T_0)$ at the initial time.
	\item
	We numerically integrate Equation~(\ref{eq.inversion}). We use the 4th order Runge-Kutta method with equally-spaced $N$ grid points in $[0, \Theta_0]$: $0 = \Theta_{N-1} < \Theta_{N-2} < \cdots < \Theta_1 < \Theta_0$. The corresponding observer times $T_\mathrm{f} = T_{N-1} > T_{N-2} > \cdots > T_1 > T_0$ are given by Equations~(\ref{eq.Theta}) and (\ref{eq.A}). The function $K(T,\theta,E(\theta))$ in Equations~(\ref{eq.inversion}) and (\ref{eq.constraint1}) is given by Equation~(\ref{eq.K}), which integrates the contribution to the observed flux in $\phi$ direction for a fixed $\theta$. The Lorentz factor of the shock wave and shocked fluid ($\Gamma_\sh$ and $\Gamma$, respectively) are described by Equations~(\ref{eq.Gammabetash}) and (\ref{eq.Gammabeta}), respectively, where the coefficient $C_\BM$ is given by Equation~(\ref{eq.C_BM}). Note that Equations~(\ref{eq.Gammabetash}) and (\ref{eq.Gammabeta}) give accurate shock and fluid speeds for relativistic regions with $\Gamma \gg 1$, which is a good approximation for the light curve segment used for inversion. This shock Lorentz factor leads to the shock radius $\Rs$ given by Equation~(\ref{eq.Rs}). The local synchrotron emissivity $\emisd$ is calculated by Equation~(\ref{eq.emisd}) with Equations~(\ref{eq.emisdpeak}), (\ref{eq.numd}), and (\ref{eq.nud}). Note that we assumed here that the observed frequency lies between the synchrotron characteristic frequency and the cooling frequency: $\numd \le \nu' \le \nucd$. The quantities that appear in Equation~(\ref{eq.K}) are evaluated at the laboratory time $\ts$ that corresponds to a given observer time $T$, where $\ts$ for each $(\theta,\phi)$ coordinate is given by Equation~(\ref{eq.ts}). $\diff K/\diff T$ in Equations~(\ref{eq.inversion}) and (\ref{eq.constraint2}) is calculated with Equations~(\ref{eq.dKdT}), (\ref{eq.dtsdT}), and (\ref{eq.dKdts}).
	\item
	As a check process after inversion, we synthesize light curves by using the reconstructed jet structure with the non-approximated original flux equation, Equation~(\ref{eq.Fbasic}). This process will be necessary because the synthesized light curve should be compared to the observed one at $T>T_\mathrm{f}$, which is the time domain that was not used for inversion. If the synthesized light curve does not match the given one, the inversion process is tried again after adjusting the parameter values.
\end{enumerate}

\section{Results}\label{sec.results}
\subsection{Demonstration of the approximated flux equation}\label{sec.results_demonstration}
Before performing inversion, we show that the approximations used for deriving Equation~(\ref{eq.inversion}) are indeed accurate, by comparing the exact flux calculated by Equation~(\ref{eq.Fbasic}) and the approximated one calculated by Equation~(\ref{eq.Fapp}), which corresponds to Equation~(\ref{eq.inversion}). 
Hereafter, we adopt $\fb = 7$, which gives a good approximation as shown below. See Appendix~\ref{app.fb} for the comparison with the other values of $\fb$.

\begin{figure}
	\includegraphics[bb = 0 0 461 461, width = \columnwidth]{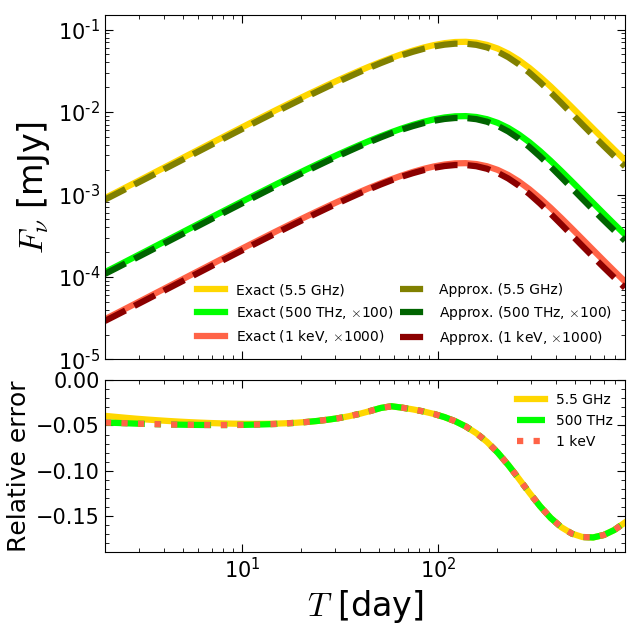}
	\caption{Upper: Synthesized light curves for radio (yellow line), optical (green line), and X-ray (red line). The solid lines are exact light curves produced by Equation~(\ref{eq.Fbasic}), which are the same as those in Figure~\ref{fig.demonstration}, while the dashed ones are approximated light curves produced by Equation~(\ref{eq.Fapp}). Lower: Relative error for each frequency, which is defined by $[F_\nu(\mathrm{Approx.}) - F_\nu(\mathrm{Exact})]/F_\nu(\mathrm{Exact})$. We can see the relative errors in the rising portion $2 \le T/\mathrm{day}\lesssim 130$ are $\lesssim 5$ per cent. The relative errors increase to $\sim 18$ per cent after the peak ($130\lesssim T/\mathrm{day}\le900$), where the shock decelerates and the relativistic approximation starts to break down.}
	\label{fig.compare}
\end{figure}

Figure~\ref{fig.compare} compares the light curves produced by Equations~(\ref{eq.Fbasic}) and (\ref{eq.Fapp}) for the Gaussian jet structure that was introduced in Section~\ref{sec.review}. The approximations used in our method is clearly justified by the similarity between the exact and approximated light curves. The relative errors in the rising phase $2 \le T/\mathrm{day} \lesssim 130$, which is used for inversion, are within $\sim5$ per cent for the three displayed frequencies.

\begin{figure*}
	\begin{tabular}{cc}
		\begin{minipage}{0.45\hsize}
			\begin{center}
				\includegraphics[bb = 0 0 384 338, width=\textwidth]{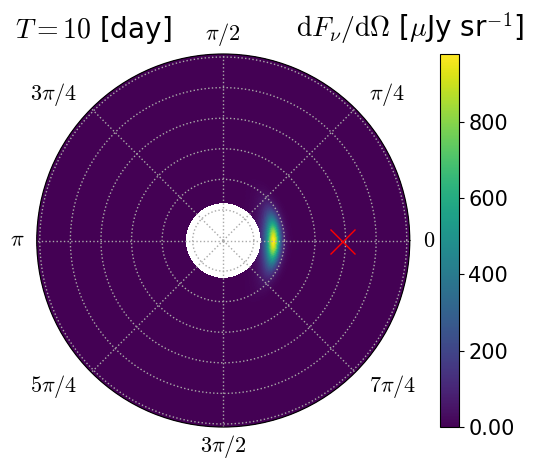}
			\end{center}
		\end{minipage} &
		\begin{minipage}{0.45\hsize}
			\begin{center}
				\includegraphics[bb = 0 0 384 338, width=\textwidth]{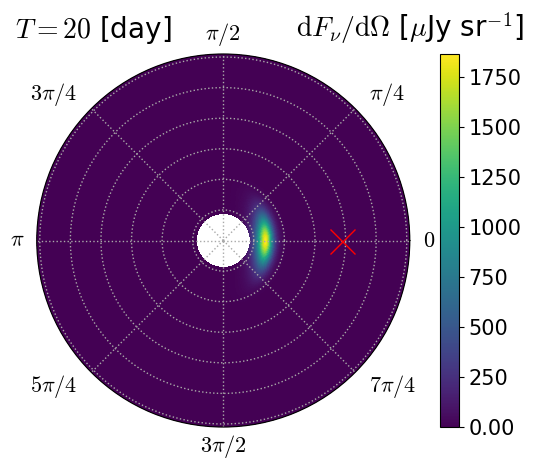}
			\end{center}
		\end{minipage} \\
		\begin{minipage}{0.45\hsize}
			\begin{center}
				\includegraphics[bb = 0 0 384 338, width=\textwidth]{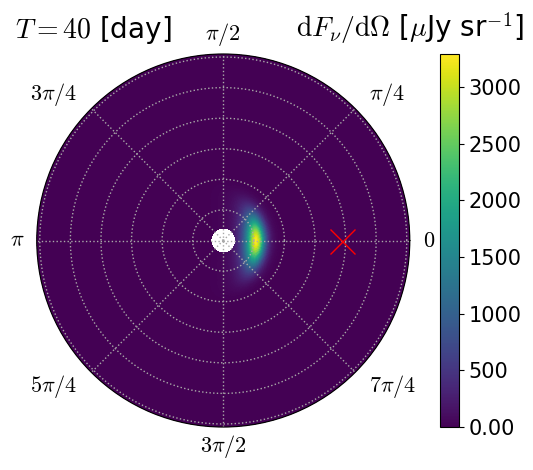}
			\end{center}
		\end{minipage} &
		\begin{minipage}{0.45\hsize}
			\begin{center}
				\includegraphics[bb = 0 0 384 338, width=\textwidth]{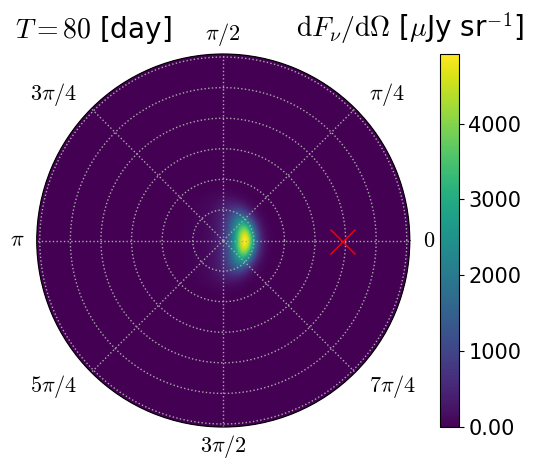}
			\end{center}
		\end{minipage} \\
		\begin{minipage}{0.45\hsize}
			\begin{center}
				\includegraphics[bb = 0 0 384 338, width=\textwidth]{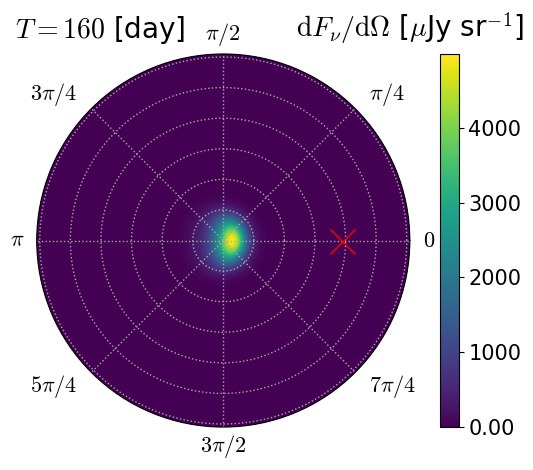}
			\end{center}
		\end{minipage} &
		\begin{minipage}{0.45\hsize}
			\begin{center}
				\includegraphics[bb = 0 0 384 338, width=\textwidth]{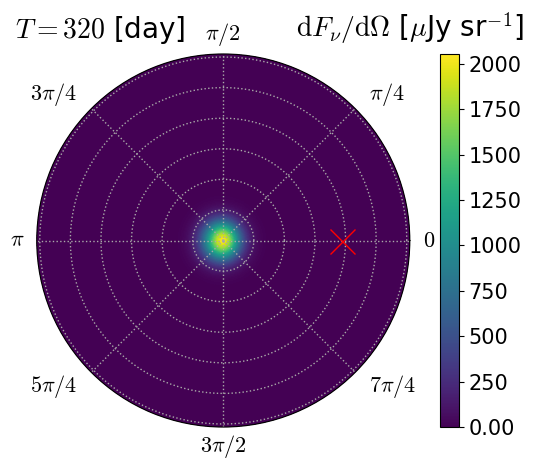}
			\end{center}
		\end{minipage} 
	\end{tabular}
	\caption{Same as Figure~\ref{fig.dist_Lv0} but $\diff F_{\nu}/\diff \Omega$ was calculated by using Equation~(\ref{eq.Fapp}) instead of Equation~(\ref{eq.Fbasic}). Unobservable regions with $\theta < \Theta(T)$ are not coloured. As seen in each panel, $\Theta(T)$ given by Equation~(\ref{eq.Theta}) traces the inner edge of the luminous region.}
	\label{fig.dist_Lv10}
\end{figure*}

We also show colour maps of $\diff F_\nu/\diff \Omega$ for the approximated radio ($\nu=5.5$~GHz) light curve in Figure~\ref{fig.dist_Lv10}, which corresponds to Figure~\ref{fig.dist_Lv0}. As evidently shown, $\Theta(T)$ well traces the inner edge of luminous regions. The colour map in the observable region $\Theta(T) \le \theta \le \thj$ in each panel is similar to that in Figure~\ref{fig.dist_Lv0}. Thus, Figure~\ref{fig.dist_Lv10} also manifests that Equation~(\ref{eq.Fapp}) is a good approximation to Equation~(\ref{eq.Fbasic}).

\begin{figure*}
	\begin{tabular}{cc}
		\begin{minipage}{0.45\hsize}
			\begin{center}
				\includegraphics[bb = 0 0 461 461, width=\textwidth]{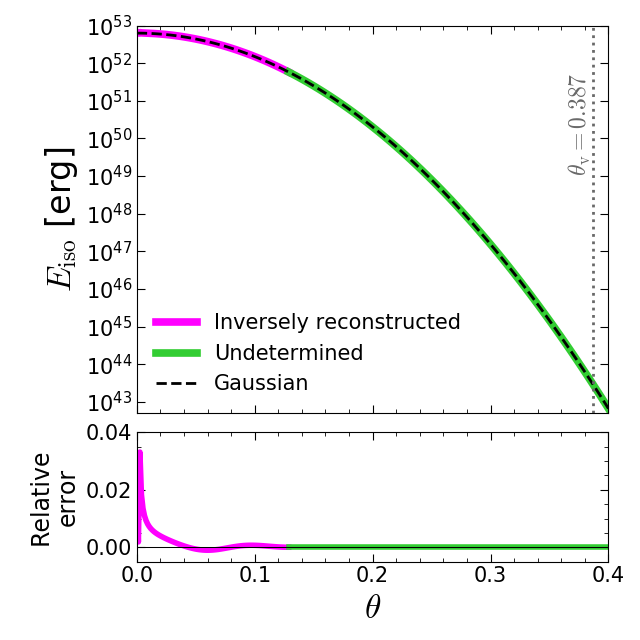}
			\end{center}
		\end{minipage} &
		\begin{minipage}{0.45\hsize}
			\begin{center}
				\includegraphics[bb = 0 0 461 461, width=\textwidth]{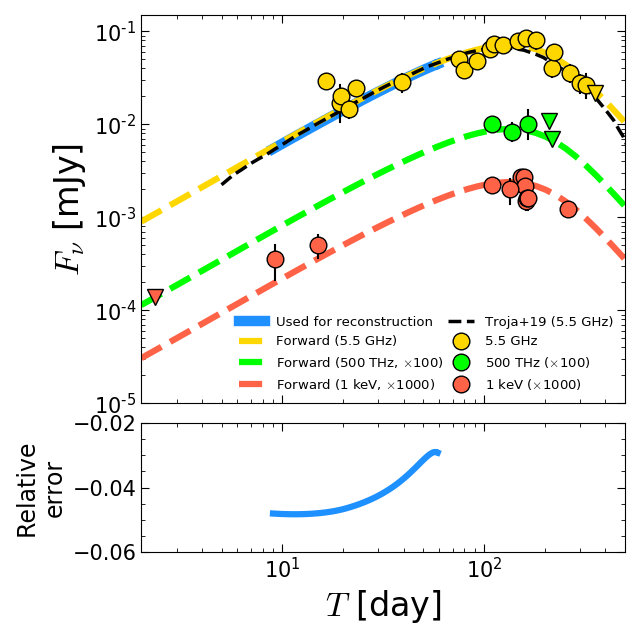}
			\end{center}
		\end{minipage} 
	\end{tabular}
	\caption{Test problem of inverse reconstruction for a Gaussian jet. Upper left: The black dashed line shows the original Gaussian given by Equation~(\ref{eq.jet_Gaussian}) with $\Ec=10^{52.8}$~erg $\sim$ $6.31\times 10^{52}$~erg, $\thc=0.059$, and $\thj=0.61$. The green line corresponds to the jet edge part that is assumed for inversion. Note that the current observations cannot in principle determine this edge part. The magenta line shows the jet structure that is inversely reconstructed. Lower left: Relative error of the reconstructed jet structure with respect to the original one, which is defined by $[E_\mathrm{iso}(\mathrm{reconstructed}) - E_\mathrm{iso}(\mathrm{original})]/E_\mathrm{iso}(\mathrm{original})$. Upper right: The blue line shows the light curve that is used for the inverse reconstruction. The yellow, green, and red dashed lines are forwardly calculated by using the inversely reconstructed jet structure that is shown in the left panel and the non-approximated equation, Equation~(\ref{eq.Fbasic}). Just for reference, we plot the observed data and the best-fitting radio light curve taken from \citet{Troja19} (black dashed line), which are the same as in Figure~\ref{fig.demonstration}. Lower right: Relative error of the radio light curve used for inversion with respect to the forwardly synthesized radio light curve, which is given by $[F_\nu(\mathrm{used}) - F_\nu(\mathrm{forward})]/F_\nu(\mathrm{forward})$.}
	\label{fig.Gaussian}
\end{figure*}

\begin{figure*}
	\begin{tabular}{cc}
		\begin{minipage}{0.45\hsize}
			\begin{center}
				\includegraphics[bb = 0 0 461 461, width=\textwidth]{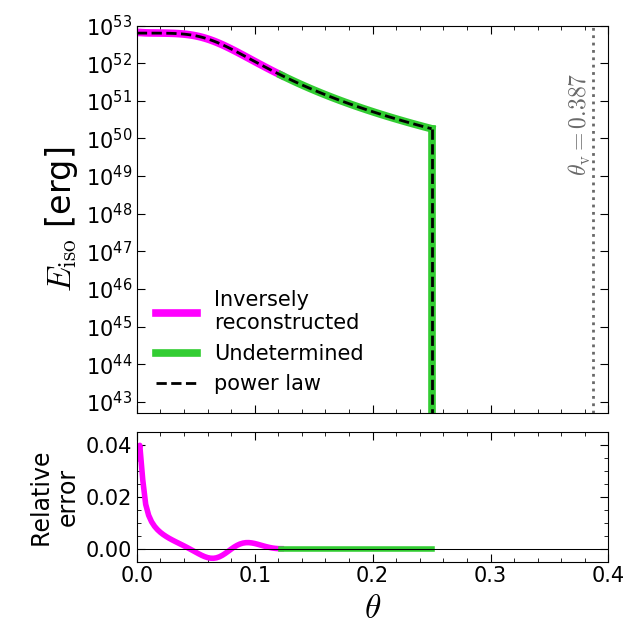}
			\end{center}
		\end{minipage} &
		\begin{minipage}{0.45\hsize}
			\begin{center}
				\includegraphics[bb = 0 0 461 461, width=\textwidth]{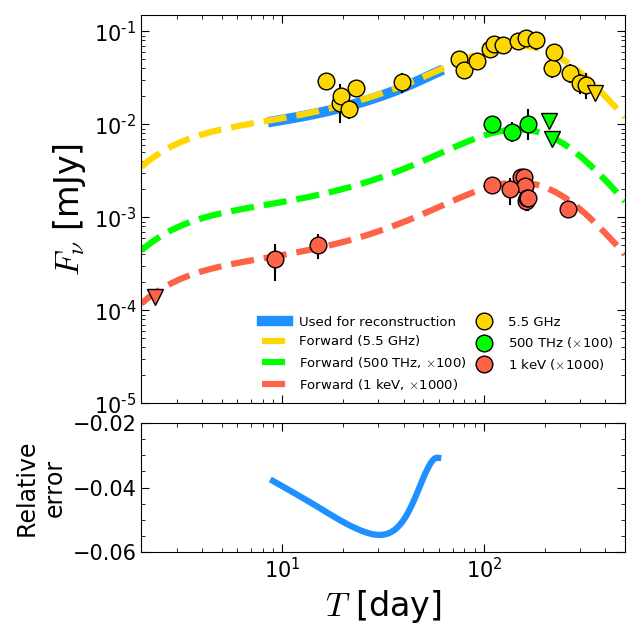}
			\end{center}
		\end{minipage} 
	\end{tabular}
	\caption{Same as Figure~\ref{fig.Gaussian} but for a power-law jet. The black dashed line in the upper left panel shows the original power-law jet given by Equation~(\ref{eq.jet_power-law}) with $\Ec=10^{52.8}$~erg $\sim$ $6.31\times 10^{52}$~erg, $\thc=0.072$, $s=4.7$, and $\thj=0.25$.}
	\label{fig.PL}
\end{figure*}

\begin{figure*}
	\includegraphics[bb = 0 0 1008 461, width = \textwidth]{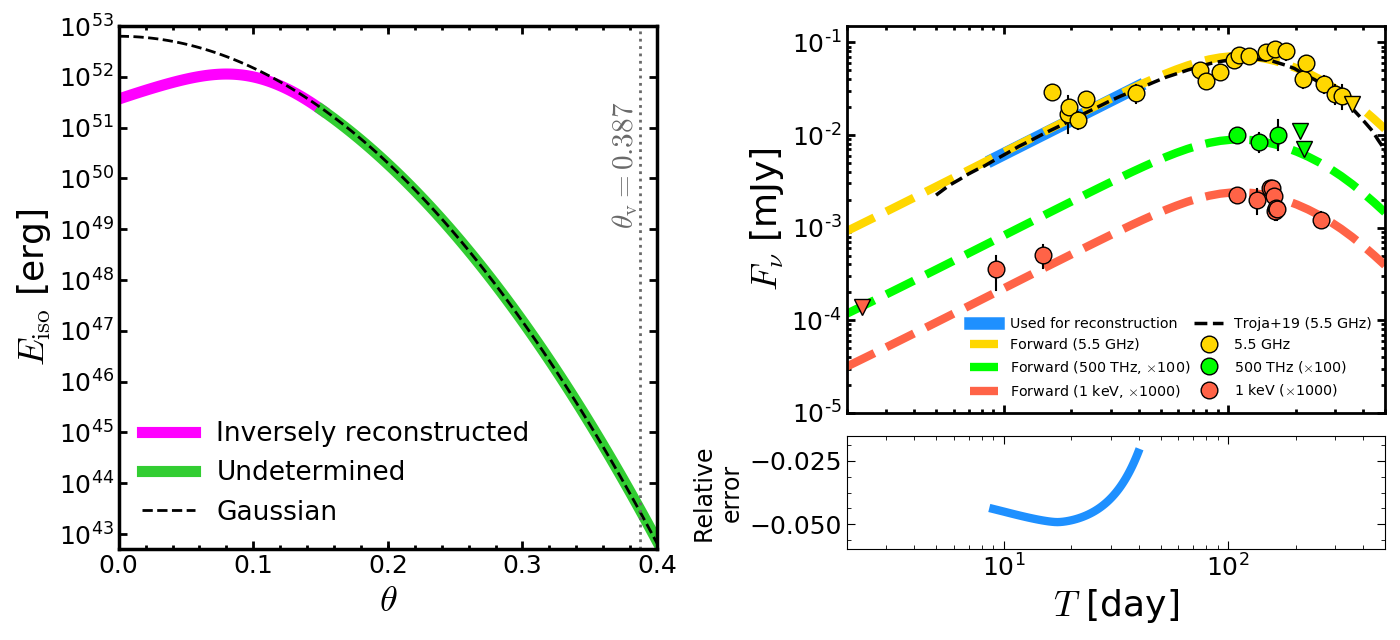}
	\caption{Inversely reconstructed hollow-cone jet. Left: The magenta line is the inversely reconstructed distribution while the green one is the distribution assumed for inversion since the current observations cannot determine this part. For reference, we draw the black dashed line that shows the Gaussian given by Equation~(\ref{eq.jet_Gaussian}) with $\Ec=10^{52.8}$~erg $\sim$ $6.31\times 10^{52}$~erg, $\thc=0.059$, and $\thj=0.61$. Upper right: The blue line shows the light curve that is used for the inverse reconstruction: $F_\nu(T) = F_{\nu,0}(T/T_0)^\alpha$ $(9\le T/\mathrm{day}\le39.7)$ with $\nu=5.5$~GHz, $F_{\nu,0}=5.45$~$\mu$Jy, $T_0=9$~days, and $\alpha=1.22$. The yellow, green, and red dashed lines are forwardly calculated by using the inversely reconstructed jet structure that is shown in the left panel and the non-approximated equation, Equation~(\ref{eq.Fbasic}). Just for reference, the observed data and the best-fitting radio light curve taken from \citet{Troja19} (black dashed line) are plotted, which are the same as in Figure~\ref{fig.demonstration}. Lower right: Relative error of the radio light curve used for inversion with respect to the forwardly calculated radio light curve, which is defined by $[F_\nu(\mathrm{used}) - F_\nu(\mathrm{forward})]/F_\nu(\mathrm{forward})$.}
	\label{fig.Cowboyhat}
\end{figure*}

\subsection{Test problems for inversion}
We demonstrate our inversion method by giving some examples. We first consider two test problems to show that our inversion formula, Equation~({\ref{eq.inversion}}), works correctly: We give a jet structure, which we call an original structure, and inversely reconstruct it from the synthesized light curve. We synthesize a light curve by using the original jet structure and Equation~(\ref{eq.Fapp}) with some fixed parameters of $\{n_0, \eB, \ee, p, \thj, \view, D\}$. Then, for inversion, we use the same parameter values of $\{n_0, \eB, \ee, p, \thj, \view, D\}$. We also give a jet structure in the jet edge part, $E(\theta, a, b)$ $(\Theta_0 \le \theta \le \thj)$, in a functional form that becomes the same as the original one if the free parameters $a$ and $b$ are correctly adjusted. For a given $T_0$, $a$ and $b$ are determined by using the synthesized light curve with Equations~(\ref{eq.constraint1}) and (\ref{eq.constraint2}), where $a$ and $b$ should turn out to be the same as the original ones in our test. Finally, the remaining structure, $E(\theta)$ $(0 \le \theta < \Theta_0)$, is inversely obtained by using Equation~(\ref{eq.inversion}), which should result in the same as the original one.

\subsubsection{Gaussian jet structure}\label{sec.results_Gaussian}
We consider a Gaussian jet described by Equation~(\ref{eq.jet_Gaussian}) with $\log (\Ec/\mathrm{erg}) = 52.8$ and $\thc = 0.059$ while we truncate the jet at $\thj=0.61$. We fix the other parameter values as $\log (n_0/\mathrm{cm}^{-3})=-2.28$, $\log \eB = -4.68$, $\log \ee = -1.39$, $p=2.17$, $\view=0.387$, and $D=41$~Mpc. We tuned these parameter values for the Gaussian jet so that the synthesized light curves become consistent with the afterglow data of GRB~170817A, as shown in the right panel of Figure~\ref{fig.Gaussian} (coloured dashed curves).\footnote{Note that the observed data points nor the best-fitting light curve taken from \citet{Troja19} in Figure~\ref{fig.Gaussian} are not directly fitted. We use them just for reference in tuning the parameters.} We choose $\nu=5.5$~GHz as the observed frequency and $T_0=9$~days as the initial time in the following inversion procedure, which leads to $F_\nu(T_0)=5.45$~$\mu$Jy and $\diff \log F_\nu/\diff \log T(T_0)=1.22$.

We assume the jet energy distribution in the jet edge part as
\begin{equation}
\label{eq.GaussianEdge}
E(\theta) = a \exp\left(-\frac{\theta^2}{2b^2}\right)\ (\Theta_0 \le \theta \le \thj),
\end{equation}
where $\thj=0.61$ is the same as for the original distribution.
Equations~(\ref{eq.constraint1}) and (\ref{eq.constraint2}) successfully recover $a=\Ec$ and $b=\thc$ in this test problem. These values give $\Theta_0=0.129$. The obtained edge structure is depicted in the left panel of Figure~\ref{fig.Gaussian} (green line).

Finally, we use Equation~(\ref{eq.inversion}) to inversely reconstruct the energy distribution. The reconstructed structure is shown in the magenta line in the left panel of Figure~\ref{fig.Gaussian}, which agrees well with the given original Gaussian structure. The portion of the light curve that is used for the inversion is shown in the blue solid line in the right panel of Figure~\ref{fig.Gaussian}, where $T_\mathrm{f}=58.9$~days is before the afterglow peak time, $T_\mathrm{p}\sim 130$~days.\footnote{The end point of the used light curve, $\Tf=58.9$~days, is slightly different from the value calculated by using the original structure and Equation~(\ref{eq.Tf}), $\Tf=58.6$~days, because of the error in the reconstructed energy at the jet axis that is shown in the lower left panel of Figure~\ref{fig.Gaussian}.}

\subsubsection{Power-law jet structure}\label{sec.results_PL}
We also check the consistency by using another jet structure. We here consider the following jet energy distribution:
\begin{equation}
\label{eq.jet_power-law}
E(\theta) = \frac{\Ec}{1 + (\theta/\thc)^s}\ (\theta \le \thj),
\end{equation}
with $s=4.7$, $\log (\Ec/\mathrm{erg}) = 52.8$, $\thc=0.072$, and $\thj=0.25$. We note that Equation~(\ref{eq.jet_power-law}) has an asymptotic form of a power law: $E\sim \Ec (\theta/\thc)^{-s}\ (\theta \gg \thc).$ Hence, the jet structure given by Equation~(\ref{eq.jet_power-law}) is often simply called a power-law jet.
The light curves synthesized by this jet structure are shown in the right panel of Figure~\ref{fig.PL} (dashed curves), where we used the same parameter values as in the previous Gaussian case: $\log (n_0/\mathrm{cm}^{-3})=-2.28$, $\log \eB = -4.68$, $\log \ee = -1.39$, $p=2.17$, $\view=0.387$, and $D=41$~Mpc. We choose $\nu=5.5$~GHz and $T_0=9$~days for inversion, which leads to $F_\nu(T_0)=10.8$~$\mu$Jy and $\diff \log F_\nu/\diff \log T(T_0)=0.362$.

We assume the following power-law structure in the jet edge part:
\begin{equation}
\label{eq.PLedge}
E(\theta) = \frac{a}{1 + (\theta/\thc)^b}\ (\theta \le \thj),
\end{equation}
where $\thc=0.072$ and $\thj=0.25$ are the same as those for the original distribution. Equations~(\ref{eq.constraint1}) and (\ref{eq.constraint2}) numerically recover $a=\Ec$ and $b=s$, which results in the jet edge structure displayed in the left panel of Figure~\ref{fig.PL} (green line).

The magenta line in the left panel of Figure~\ref{fig.PL} shows the inversely reconstructed energy distribution obtained by Equation~(\ref{eq.inversion}). As shown in the figure, our method successfully reconstructs the original power-law jet structure. The light curve that is used in the inversion process is shown in the right panel of Figure~\ref{fig.PL} (blue solid line), where $T_\mathrm{f}= 59.3$~days, 
which is before the afterglow peak.\footnote{Same as in the Gaussian case, the end point of the used light curve, $\Tf=59.3$~days, is different from the value calculated by using the original structure and Equation~(\ref{eq.Tf}), $\Tf=58.6$~days, because of the error in the reconstruction shown in the lower left panel of Figure~\ref{fig.PL}.} Note that the observed data points in Figure~\ref{fig.PL} are not used for fitting but plotted just for reference.

As shown in the above examples, our inversion formula works very well. In the next subsection, we consider a more practical problem.

\subsection{An example of inversion: Hollow-cone jet structure} \label{sec.hollow}
This subsection presents a more practical example. We prepare by hand a light curve that agrees with the observed data points of the afterglow of GRB~170817A. As one of the simplest examples, we assume 
a light curve with a constant slope in the log-log plane given by
\begin{equation}
\label{eq.hollowLC}
F_\nu (T) = F_{\nu,0}\left(\frac{T}{T_0}\right)^\alpha,
\end{equation}
where $\nu = 5.5$~GHz, $\alpha = \diff \log F_\nu/\diff \log T$ is a constant, and $F_{\nu,0}$ is the flux density observed at $T=T_0$. We choose $\alpha = 1.22$ to roughly connect the rising part of the radio observational data,
and set $T_0=9$~days and $F_\nu(T_0)=F_{\nu,0}=5.45$~$\mu$Jy. Note that $T_0$, $F_\nu(T_0)$, and $\diff \log F_\nu/\diff \log T(T_0)$ are the same as in the example in Section~\ref{sec.results_Gaussian}.

The jet edge part is assumed to be the Gaussian described by Equation~(\ref{eq.GaussianEdge}) with the jet truncation angle $\thj=0.61$. We adopt $\log (n_0/\mathrm{cm}^{-3})=-3.01$ and $\log \eB = -3.56$, which are tuned to adjust the peak time and peak flux of the light curves that are synthesized with the reconstructed structure, while the other parameter values are the same as those in Section~\ref{sec.results_Gaussian}: $\log \ee = -1.39$, $p=2.17$, $\view=0.387$, and $D=41$~Mpc. The free parameters $a$ and $b$ are determined by Equations~(\ref{eq.constraint1}) and (\ref{eq.constraint2}) as $a=52.8$~erg and $b=0.0593$, which give $\Theta_0=0.151$. The obtained edge structure is shown in the left panel of Figure~\ref{fig.Cowboyhat} (green line), which is almost indistinguishable from the Gaussian in Section~\ref{sec.results_Gaussian} (dashed line).

By using the above setup, we inversely reconstruct the jet structure with Equation~(\ref{eq.inversion}). The obtained structure is shown in the left panel of Figure~\ref{fig.Cowboyhat} (magenta line). Interestingly, the inversely reconstructed jet structure is non-trivial and a so-called hollow-cone structure, not a Gaussian nor a power law. The jet energy peaks around $\theta \sim 0.08 \sim 4.6$~deg and the jet axis has lower energy, which is about an order of magnitude smaller than that in the previous Gaussian example (dashed line). The portion of the light curve used for the inversion is shown in the right panel of Figure~\ref{fig.Cowboyhat} (blue solid line), where $T_\mathrm{f}=39.7$~days. This jet structure synthesizes the light curves that agree well with the whole data points of the afterglow as shown in the panel (coloured dashed curves), whereas the observed data points are not directly used for fitting but plotted just for reference.

The reason why the constant slope light curve $F_\nu \propto T^{1.22}$ leads to a hollow-cone jet is explained as follows in comparison with the case of the Gaussian jet in Section~\ref{sec.results_Gaussian}. The light curve produced by a Gaussian jet is convex upward as shown in Figure~\ref{fig.Gaussian}. That is, the increasing rate of the observed flux is smaller than that of the constant slope light curve, because higher energy in the inner region (as in the Gaussian jet) leads to more delayed contribution to the afterglow emission due to the relativistic beaming. To keep the increasing rate constant, $\diff \log F_\nu/\diff \log T =1.22$, the inner region has to be visible earlier and contribute to the observed flux, which requires the hollow-cone jet structure with lower jet energy than that for the Gaussian jet.

\begin{figure*}
	\begin{tabular}{cc}
		\begin{minipage}{0.45\hsize}
			\begin{center}
				\includegraphics[bb = 0 0 461 346, width=\textwidth]{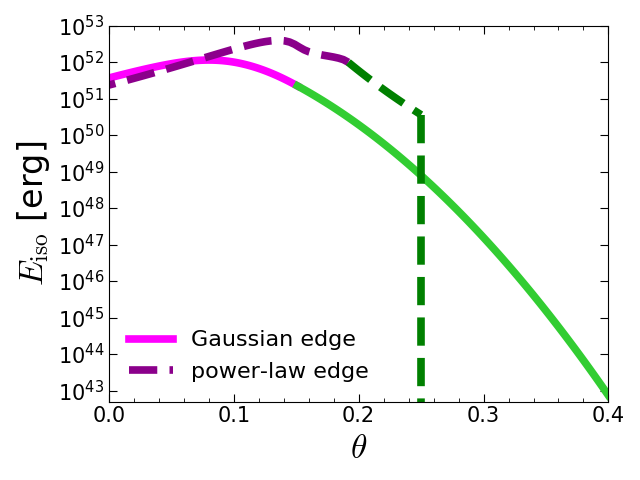}
			\end{center}
		\end{minipage} &
		\begin{minipage}{0.45\hsize}
			\begin{center}
				\includegraphics[bb = 0 0 461 461, width=\textwidth]{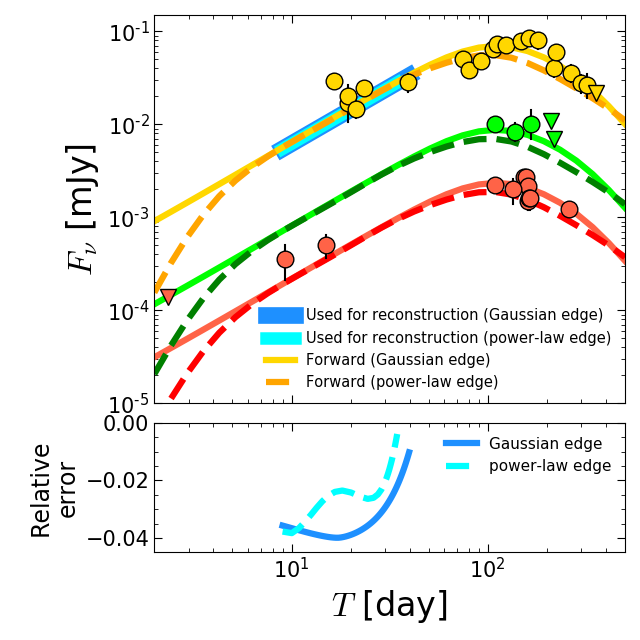}
			\end{center}
		\end{minipage}
	\end{tabular}
	\caption{Dependence on the edge structure. Left: Reconstructed jet structures for a Gaussian edge (solid), which is the same as in Figure~\ref{fig.Cowboyhat}, and a power-law edge (dashed). The magenta lines show the reconstructed structure while the green lines show the assumed edge structure. Upper right: Forwardly synthesized light curves (yellow lines for $\nu=5.5$~GHz, green for $\nu=500$~THz, and red for 1~keV) and radio light curves used for inversion (blue ones). The solid lines correspond to the jet with the Gaussian edge while the dashed ones correspond to the jet with the power-law edge. The observed data points are plotted just for reference. Lower right: Relative error between the forwardly synthesized light curve and the used one for inversion, which is defined by $[F_\nu(\mathrm{used}) - F_\nu(\mathrm{forward})]/F_\nu(\mathrm{forward})$.}
	\label{fig.edge}
\end{figure*}

\begin{figure*}
	\begin{tabular}{cc}
		\begin{minipage}{0.45\hsize}
			\begin{center}
				\includegraphics[bb = 0 0 461 461, width=\textwidth]{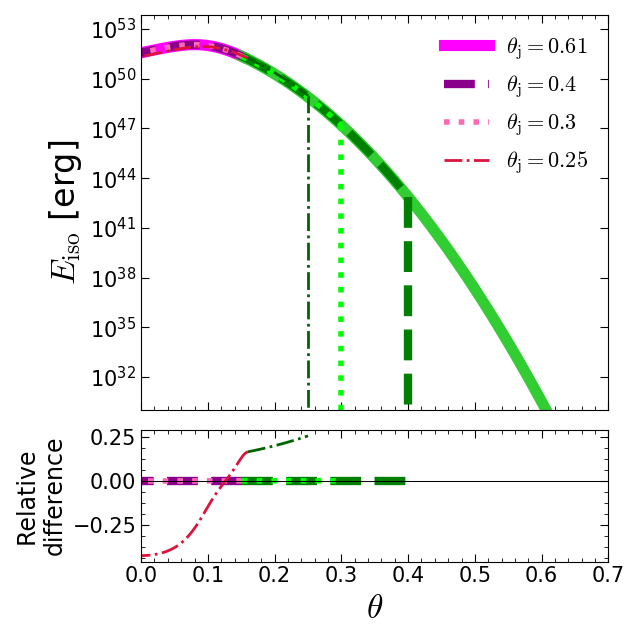}
			\end{center}
		\end{minipage} &
		\begin{minipage}{0.45\hsize}
			\begin{center}
				\includegraphics[bb = 0 0 461 461, width=\textwidth]{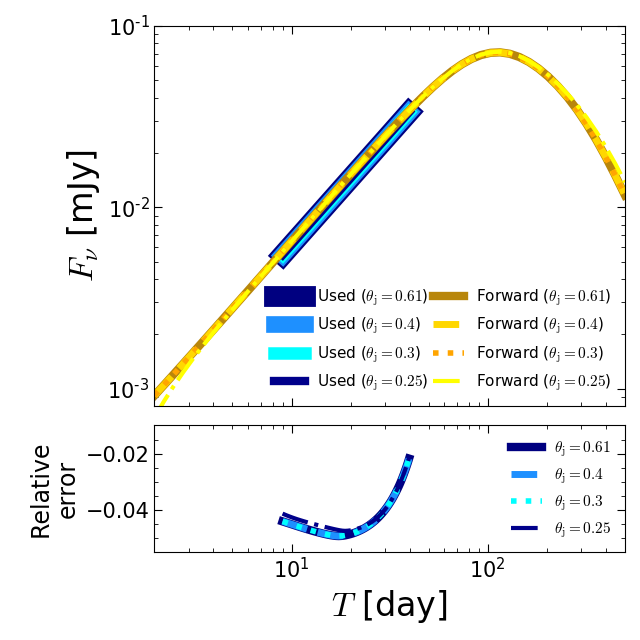}
			\end{center}
		\end{minipage}
	\end{tabular}
	\caption{Dependence on the jet truncation angle $\thj$. Upper left: Reconstructed jet structures with different $\thj$ ($\thj=0.61$: solid line, $\thj=0.4$: dashed one, $\thj=0.3$: dotted one, $\thj=0.25$: dot-dashed one). Magenta lines show the reconstructed portion while the green ones present the assumed edge structure. Lower left: Relative difference of each jet structure with respect to the fiducial hollow-cone jet obtained for $\thj=0.61$, which is defined by $[E_\mathrm{iso}(\mathrm{reconstructed}) - E_\mathrm{iso}(\mathrm{fiducial})]/E_\mathrm{iso}(\mathrm{fiducial})$. We can see that the reconstructed structure does not depend on the edge part so much for $\thj \ge 0.3$. Upper right: Forwardly synthesized light curves ($\nu=5.5$~GHz, yellow lines) and radio light curves used for inversion (blue ones). Lower right: Relative error between the forwardly synthesized light curve and the used one for inversion, which is defined by $[F_\nu(\mathrm{used}) - F_\nu(\mathrm{forward})]/F_\nu(\mathrm{forward})$.}
	\label{fig.thj}
\end{figure*}

\begin{figure*}
	\begin{tabular}{cc}
		\begin{minipage}{0.45\hsize}
			\begin{center}
				\includegraphics[bb = 0 0 461 346, width=\textwidth]{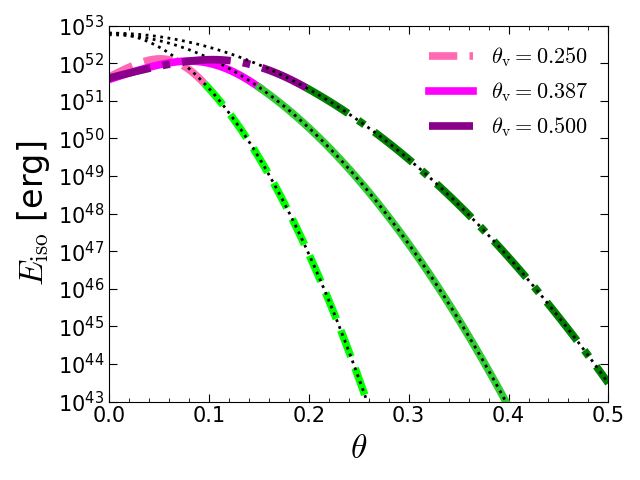}
			\end{center}
		\end{minipage} &
		\begin{minipage}{0.45\hsize}
			\begin{center}
				\includegraphics[bb = 0 0 461 461, width=\textwidth]{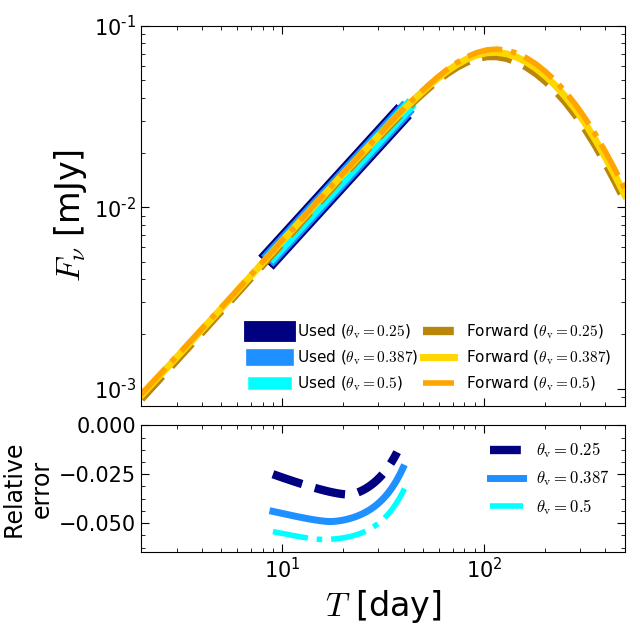}
			\end{center}
		\end{minipage}
	\end{tabular}
	\caption{Dependence on the viewing angle $\view$. Left: Reconstructed jet structures for $\view=0.25$ (dashed), $0.338$ (solid), and $0.5$ (dash-dotted). The magenta lines show the reconstructed structure while the green lines show the assumed edge structure. For reference, we plot Gaussian shapes by black dotted lines. Upper right: Forwardly synthesized light curves ($\nu=5.5$~GHz, yellow lines) and radio light curves used for inversion (blue ones). Lower right: Relative error between the forwardly synthesized light curve and the used one for inversion, which is defined by $[F_\nu(\mathrm{used}) - F_\nu(\mathrm{forward})]/F_\nu(\mathrm{forward})$.}
	\label{fig.view}
\end{figure*}

\subsection{Parameter dependence}
Here, we study the parameter dependence of the reconstructed jet structure by using the same light curve as in Section~\ref{sec.hollow}, which is given by Equation~(\ref{eq.hollowLC}) with $\nu = 5.5$~GHz, $\alpha=1.22$, $T_0=9$~days, and $F_\nu(T_0)=F_{\nu,0}=5.45$~$\mu$Jy. As shown below, the hollow-cone structure is always reconstructed whereas each jet structure is quantitatively different. The physical reason has been discussed in the last part of Section~\ref{sec.hollow}.

First, we investigate the dependence on the edge structure by changing the functional form. For comparison with the Gaussian edge, we use the same power-law edge as in Section~\ref{sec.results_PL}, which is given by Equation~(\ref{eq.PLedge}) with $\thc=0.072$ and $\thj=0.25$, while the free parameters $a$ and $b$ are newly determined. We tune $n_0$ and $\eB$ for adjusting the peak time and peak flux of the forwardly synthesized light curve while keeping the other parameters the same as in Section~\ref{sec.hollow}. Figure~\ref{fig.edge} shows the reconstructed jet structure and the corresponding light curves for the power-law edge (dashed lines) in comparison with the Gaussian edge (solid ones). The adjusted parameters are $\log (n_0/\mathrm{cm}^{-3})=-3.02$ and $\log \eB=-4.68$ for the case of the power-law edge, which leads to $a=10^{57.3}$~erg and $b=12.5$. The reconstructed jet has another hollow-cone structure and the synthesized light curves are consistent with the observations.

We then check the dependence on the jet truncation angle, $\thj$. Figure~\ref{fig.thj} shows the jet structure obtained for $\thj=0.61$, $0.4$, $0.3$, and $0.25$ and the corresponding light curves, where we used the same parameter values as in Section~\ref{sec.hollow} except for $\thj = 0.25$, for which we employed $\log (n_0/\mathrm{cm}^{-3}) = -3.25$ and $\log \eB = -3.25$ to adjust the peak time and peak flux of the forwardly synthesized light curve. As shown in the left panel of Figure~\ref{fig.thj}, the reconstructed jet structures and the produced light curves are almost the same for $\thj \ge 0.3$, because the jet edge part, $0.3 \le \theta \le 0.61$, does not much contribute to the observed flux for $T\ge9$~days. For $\thj=0.25$, on the other hand, the reconstructed structure is qualitatively the same as those for $\thj \ge 0.3$ but quantitatively different from them as shown in the lower left panel of Figure~\ref{fig.thj}. The contribution from $0.25 \le \thj \le 0.3$ is lost, which is not negligible for $T\gtrsim9$~days since $\thj$ is much closer to $\Theta_0$. To compensate the lost flux, the Gaussian edge structure given by Equation~(\ref{eq.GaussianEdge}) slightly changes as indicated by the green line in the lower panel of Figure~\ref{fig.thj}. The difference of the jet edge structure cumulatively affects the inversion of the inner jet structure and the relative difference eventually increases to $\sim 40$ per cent at the jet axis.

Lastly, we study the dependence on the viewing angle $\view$. We try $\view = 0.25\sim14.3^\circ$ and $\view=0.5\sim28.6^\circ$, which are respectively the smallest and largest viewing angles of GRB~170817A inferred from the superluminal motion \citep{superluminal}. We only tune $n_0$ and $\eB$ for each $\view$ to adjust the peak time and peak flux while keeping the other parameters the same as in Section~\ref{sec.hollow}. The adjusted parameters are $\log (n_0/\mathrm{cm}^{-3})=-4.37$ and $\log \eB=-2.93$ for $\view=0.25$ while $\log (n_0/\mathrm{cm}^{-3})=-2.15$ and $\log \eB=-4.06$ for $\view=0.5$. These parameters give $a$ and $b$ in Equation~(\ref{eq.GaussianEdge}) as follows: $a=10^{52.8}$~erg and $b=0.0385$ for $\view=0.25$; $a=10^{52.8}$~erg and $b=0.0763$ for $\view=0.5$. Figure~\ref{fig.view} shows the reconstructed jet structures and corresponding radio light curves. As shown, they are hollow-cone type structures while the width of the jet becomes wider for larger $\view$. These jet structures synthesize light curves consistent with the observations.

\section{Summary \& Discussions}\label{sec.conclusion}

We formulate an inversion method that reconstructs jet structure from off-axis GRB afterglows without assuming any functional form of the structure. Based on the standard theory of GRB afterglows, we derive an ordinary differential equation, Equation~(\ref{eq.inversion}), which uniquely determines a jet structure for a given light curve and a given parameter set.
We demonstrate that the inversion formula successfully reconstructs the jet structure for a Gaussian and a power-law jet in Sections~\ref{sec.results_Gaussian} and \ref{sec.results_PL}, respectively.

The advantage of our method is that it can reconstruct non-trivial jet structures without assuming a functional form.
This is sharply contrast to the previous methods given by \citet{GG18,Ryan}, which assume Gaussian or power-law jet structures, while their analytical methods are easy to use and complementary.
Especially, our basic equation (\ref{eq.Fbasic}) is essentially the same as that in \citet{Ryan} and, overall, our results are consistent with their results, except for the central region of the jet.

Our inversion method discovers that the jet of GRB~170817A could have a hollow-cone structure as well as Gaussian and power-law structures,
given the uncertainty of the observed light curves.
The hollow-cone type of jet structure was not possible to identify by the previous methods assuming a functional form of the jet structure, and hence
has not been discussed for GRB~170817A.

There are several possibilities for the formation of hollow-cone jets.
The first possibility is that the jet is launched by the Blandford-Znajek mechanism and the Poynting flux is zero at the jet axis \citep{BZ,McKinney06,TMN08}, although it depends on the magnetic field configuration and the propagation under the cocoon pressure \citep{Kathirgamaraju19}.
The second possibility is that such structure may be formed via the interaction between the jet and the ambient medium during the propagation through the ejecta and/or at the jet breakout \citep{Z03,MI13}.
Jet precession might also produce a hollow-cone jet in the case of black hole-neutron star binary mergers with misalignment between the orbital angular momentum and the black-hole spin \citep{McKinney13,Kawaguchi15,Huang19}.
We also note that the pulsar beam structure is discussed to be a hollow-cone jet \citep{pb1, pb2}.
Some of these mechanisms may be responsible for the formation of a hollow-cone structure, whereas it is beyond the scope of this paper to pin down the formation mechanism of the jet structure.

It is still possible that the jet of GRB~170817A has a Gaussian or a power-law structure because they also synthesize the light curves consistent with the observed data.
Since the reconstructed jet structure, in particular the central part, is sensitive to the given light curve,
precise observations with high cadence are necessary for constraining the jet structure.

We also emphasize that the current observations determine
the jet structure only around the jet axis
and do not constrain
the outer jet structure around the line-of-sight viewing angle at all.
Early observations are required to determine the jet structure
near the line of sight.
Without early observations, the jet structure has huge uncertainties
at large polar angles. Note that this outer part crucially affects the detection rate of off-axis GRBs \citep{Beniamini19,IN19}.

Since this is a proof-of-concept paper,
we only consider the simplest case and find a hollow-cone jet as a new type of the jet structure.
Changing the model parameters, we find that the simple power-law light curves given by $F_\nu \propto T^{1.22}$ always reconstruct hollow-cone type jets. We need systematic surveys of possible jet structures under the uncertainties of the light curves, which will be done in a forthcoming paper.
It is also an interesting future study to consider the effects of the non-uniform ambient medium or additional energy injection from the central engine after the jet launch on the inverse reconstruction of the jet structure.

Our inversion method would be applied not only to GRB~170817A but also to other off-axis GRBs that will be detected in future, provided the viewing angle is larger than the minimal value given by Equation~(\ref{eq.off-axis}). On the other hand, the viewing angle should not be too large, since the emission from the counter jet can contaminate the afterglow light curves, which is also an interesting issue to study in a forthcoming paper.

\section*{Acknowledgements}
We thank Yizhong Fan, Dimitrios Giannios, Jonathan Granot, Amir Levinson, Takashi Nakamura, Tsvi Piran, and
Hendrik Jan van Eerten for useful discussions.
We also thank Hamid Hamidani, Wataru Ishizaki, Koutarou Kyutoku, Tatsuya Matsumoto, and Tomoki Wada for daily discussions. 
We thank the anonymous referee for the useful comments.
This work is supported by JSPS Grants-in-Aid for Scientific Research
17H06362 (KT, KI)
and 18H01213, 18H01215, 17H06357, 17H06131 (KI).

\section*{Data Availability}
The data underlying this article will be shared on reasonable request to the corresponding author.







\appendix

\section{Approximated shock radius and laboratory time}
In this section we derive an approximated shock radius $R=\Rs$ for Equation~(\ref{eq.R}) and an approximated laboratory time $t=\ts$ for Equation~(\ref{eq.t}), which are introduced to considerably reduce the computational time for inversion. The point is again that only a portion of the relativistic region contributes to the observed flux during the inverse reconstruction.

We approximate shock dynamics by simply neglecting the Sedov-Taylor term in Equations~(\ref{eq.BM-STsh}) and (\ref{eq.BM-ST}) as follows:
\begin{align}
\label{eq.Gammabetash}
\Gamma_\sh^2 \beta_\sh^2 &= C_\BM^2 t^{-3},\\
\label{eq.Gammabeta}
\Gamma^2 \beta^2 &= \frac{1}{2}C_\BM^2 t^{-3}.
\end{align}
Note that this is not the Blandford-McKee solution but approaches it in the limit of $\beta,\beta_\sh \rightarrow 1$, since $\beta$ and $\beta_\sh$ remain in the left-hand side, which keeps $\Gamma _\sh$ and $\Gamma$ above unity for any $t$. In this case, the shock radius given by Equation~(\ref{eq.R}) has the following analytic form:
\begin{equation}
\label{eq.Rs}
R = \Rs := {}_2F_1\left(\frac{1}{3},\frac{1}{2},\frac{4}{3};-\frac{t^3}{C_\BM^2}\right) ct,
\end{equation}
where ${}_2F_1(\cdots)$ is the Gauss's hypergeometric function. Substituting Equation~(\ref{eq.Rs}) to Equation~(\ref{eq.t}) and using the relativistic limit: $\Rs \rightarrow ct[1 - t^3/(8C_\BM^2)]$ as $\Gamma_\sh^2 \beta_\sh^2 = C_\BM^2t^{-3} \rightarrow \infty$, we obtain an algebraic equation for $t$:
\begin{equation}
\label{eq.quaritc}
\mu t^4 + 8C_\BM^2(1-\mu) t - 8C_\BM^2T = 0.
\end{equation}
The appropriate solution for $0\le \mu \le 1$ is given by
\begin{equation}
\label{eq.ts}
t = \ts := \left\{
\begin{array}{ll}
\displaystyle  T & (\mu = 0)\\
\displaystyle  (8C_\BM^2 T)^{1/4}& (\mu = 1) \\
\displaystyle  \sqrt{\frac{\sqrt{2}a}{Y} - \frac{Y}{2}} -\sqrt{\frac{Y}{2}} & (0<\mu<1)
\end{array} \right.,
\end{equation}
where
\begin{align}
Y &:= \frac{3^{1/3}X^2 - 3^{2/3}b}{3X},\\
X &:= [\sqrt{3(27a^4 + b^3)} + 9a^2]^{1/3},\\
a &:= \frac{2C_\BM^2(1-\mu)}{\mu},\\
\label{eq.b}
b &:= \frac{8C_\BM^2T}{\mu}.
\end{align}
The derivative of $\ts$ with respect to $T$ is given by
\begin{equation}
\label{eq.dtsdT}
\frac{\diff \ts}{\diff T} = \frac{b}{4T(\ts^3 + a)}.
\end{equation}
Equations~(\ref{eq.Rs}) and (\ref{eq.ts}) give sufficiently accurate solutions in relativistic regions as demonstrated in Section~\ref{sec.results_demonstration}.

\section{Explicit form of $\diff K/\diff \ts$}
Using Equations~(\ref{eq.emisdpeak})-(\ref{eq.numd}), (\ref{eq.emisd}), and (\ref{eq.Gammabetash})-(\ref{eq.Rs}), we obtain the derivative of Equation~(\ref{eq.K}) as follows:
\begin{align}
\label{eq.dKdts}
&\frac{\diff K}{\diff \ts} \nonumber \\
&= \frac{1}{4\pi D^2}\int _0^{2\pi} \diff \phi \frac{\sin \theta \Rs ^3 \emisd}{4\ts \Gamma^4(1-\beta \mu)^3(1-\beta_\sh \mu)^2} \nonumber \\ 
&\quad \times \left\{(1-\beta \mu)(1 -\beta _\sh \mu)\left\{2\beta^2 + \frac{c\beta_\sh \ts}{\Rs} 
- \frac{1}{4\Gamma^2}\biggl[(\Gamma+1)(4\Gamma-3)
\right. \right. \nonumber \\ 
&\quad \left. \left. + (p-1)\left(\frac{(\Gamma + 1)(6\Gamma -1)}{2} + \frac{1 - \Gamma^2(1-\beta \mu)}{1-\beta \mu}\right)\right] \right\} \nonumber \\
&\quad \left. \left. - \frac{\beta \mu (1-\beta_\sh \mu)}{\Gamma^2} - \frac{\beta_\sh \mu (1-\beta \mu)}{2\Gamma_\sh^2} \right\}\right|_{t=\ts}.
\end{align}

\section{Dependence on $\fb$} \label{app.fb}
In this section, we investigate the dependence on $\fb$, which has been fixed to $\fb=7$ so far. Shortly speaking, larger $\fb$ gives more accurate approximated light curves while reducing the reconstructed part of the jet. Since there is a trade-off between the accuracy and the extension, the reasonable value of $\fb$ depends on the purpose. In this paper, we adopt $\fb=7$ as a reasonable choice.

\begin{figure*}
	\includegraphics[bb = 0 0 701 601, width = \textwidth]{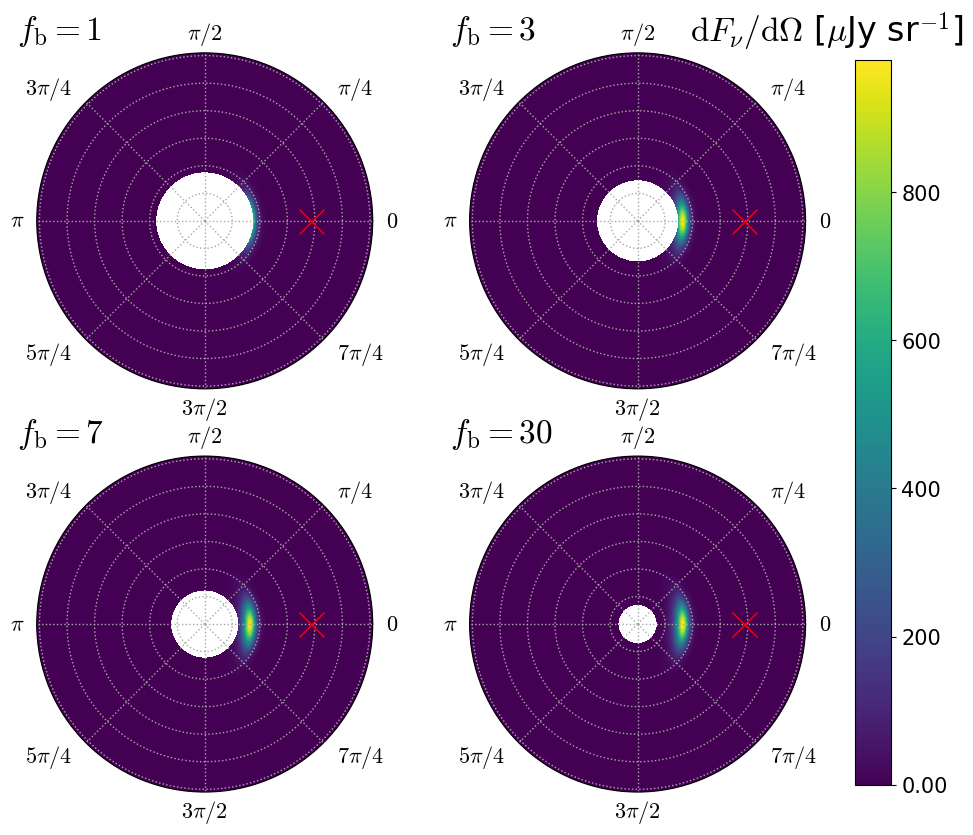}
	\caption{Same as Figure~\ref{fig.dist_Lv10} but for different $\fb$, which is displayed at the upper left corner of each panel, at a given observer time $T=10$~day.}
	\label{fig.fbdiffs}
\end{figure*}

\begin{figure*}
	\begin{tabular}{cc}
		\begin{minipage}{0.45\hsize}
			\begin{center}
				\includegraphics[bb = 0 0 461 461, width=\textwidth]{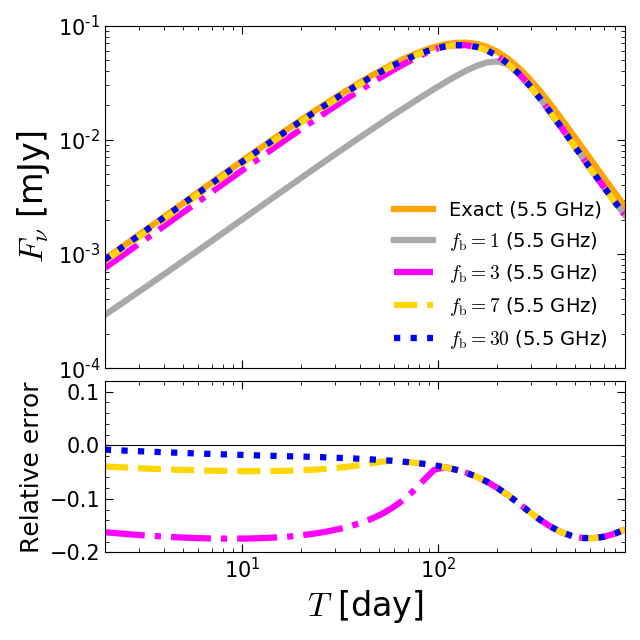}
			\end{center}
		\end{minipage} &
		\begin{minipage}{0.45\hsize}
			\begin{center}
				\includegraphics[bb = 0 0 461 346, width=\textwidth]{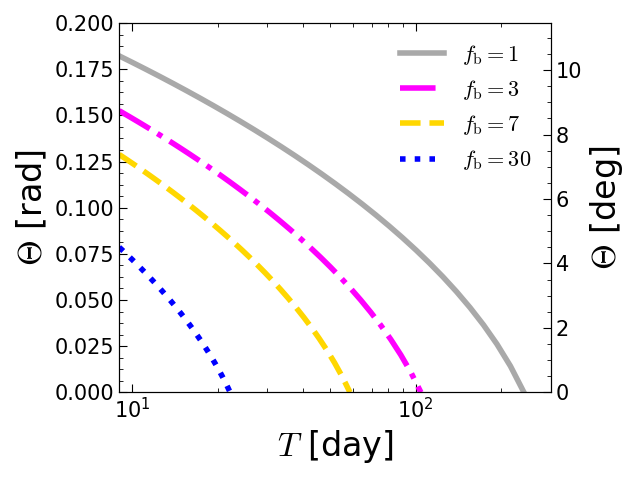}
			\end{center}
		\end{minipage}
	\end{tabular}
	\caption{Upper left: Radio ($\nu=5.5$~GHz) light curve produced by Equation~(\ref{eq.Fbasic}), which is labelled as `Exact', and those produced by Equation~(\ref{eq.Fapp}) with different values of $\fb$. Note that the exact light curve and the approximated light curve for $\fb=7$ are the same as in Figure~\ref{fig.compare}. Lower left: Relative error of the approximated light curves for $\fb=3$, $7$, and $30$ with respect to the exact one. Right: Inner truncation angle $\Theta$ as a function of the observer time $T$ for different values of $\fb$.}
	\label{fig.fbdiffs2}
\end{figure*}

Figure~\ref{fig.fbdiffs} shows the colour maps of $\diff F_\nu/\diff \Omega$ at $T=10$~day for the Gaussian jet with the parameter values that are used in Section~\ref{sec.results_demonstration}, where the observed flux is calculated by Equation~(\ref{eq.Fapp}). As stated in Section~\ref{sec.Inv-inv}, larger $\fb$ gives smaller inner cutoff $\Theta(T)$. We emphasize that if the beaming cone has a usual size, $\fb=1$, $\Theta(T)$ lies on the most luminous area and, hence, cuts the large portion of the luminous region. As a result, the approximated light curve for $\fb=1$ is significantly dimmer than the exact one as shown in the left panel of Figure~\ref{fig.fbdiffs2}. This is the essential reason why we introduce $\fb > 1$.

Since larger $\fb$ cuts less jet region, the synthesized light curve becomes more accurate for larger $\fb$ until $\Theta(T)$ becomes zero at $T = \Tf$, as shown in the left panel of Figure~\ref{fig.fbdiffs2}. However, larger $\fb$ leads to larger jet edge region that should be assumed ($\Theta(T_0) \le \theta \le \thj$), since $\Theta(T_0)$ is reduced for a given initial time $T_0$. In other words, larger $\fb$ reduces the jet inner region that is reconstructed, which lies in $\theta \le \Theta(T_0)$. It is also worth noting that $\Theta=0$ is realized earlier (i.e., $\Tf$ becomes smaller) for larger $\fb$ as shown in the right panel of Figure~\ref{fig.fbdiffs2}. As a result, larger $\fb$ reduces the portion of the light curve that is used for inversion, which is given by $T\le \Tf$. Hence, it is necessary to choose a reasonable value of $\fb$, for which the light curve is accurate enough, the reconstructed jet region is reasonably wide (i.e., $\Theta(T)$ is reasonably large), and the portion of the light curve used for inversion is not so short (i.e., $\Tf$ is not so small). The appropriate value of $\fb$ must depend on the situation. We find $\fb=7$ works well in this paper, which gives approximated light curves with relative errors around $\lesssim5$ per cent and the wide reconstructed region that is enough for the non-trivial hollow-cone structure to appear (See Figures~\ref{fig.Cowboyhat}, \ref{fig.edge}, \ref{fig.thj}, and \ref{fig.view}).

\begin{figure*}
	\begin{tabular}{cc}
		\begin{minipage}{0.45\hsize}
			\begin{center}
				\includegraphics[bb = 0 0 461 461, width=\textwidth]{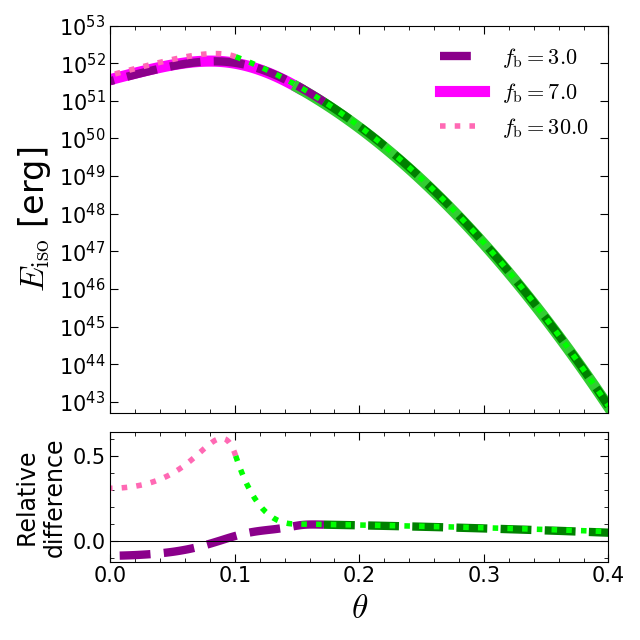}
			\end{center}
		\end{minipage} &
		\begin{minipage}{0.45\hsize}
			\begin{center}
				\includegraphics[bb = 0 0 461 461, width=\textwidth]{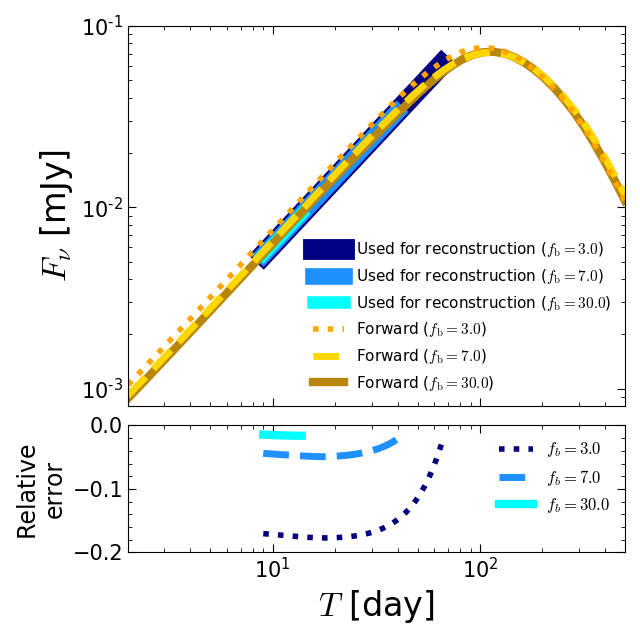}
			\end{center}
		\end{minipage}
	\end{tabular}
	\caption{Dependence on $\fb$. Upper left: Reconstructed jet structures for $\fb=3$, $7$ and $30$. Note that the jet structure for $\fb=7$ is the same as in Figure~\ref{fig.Cowboyhat}. Lower left: Relative difference with respect to the fiducial case with $\fb=7$, which is defined by $[E_\mathrm{iso}(\mathrm{reconstructed}) - E_\mathrm{iso}(\mathrm{fiducial})]/E_\mathrm{iso}(\mathrm{fiducial})$. Upper right: Forwardly synthesized light curves ($\nu=5.5$~GHz, yellow lines) and radio light curves used for inversion (blue ones). Note that the lines for $\fb=7$ are the same as in Figure~\ref{fig.Cowboyhat}. Lower right: Relative error between the forwardly synthesized light curve and the used one for inversion, which is defined by $[F_\nu(\mathrm{used}) - F_\nu(\mathrm{forward})]/F_\nu(\mathrm{forward})$.}
	\label{fig.fb}
\end{figure*}

Finally, we repeat the jet reconstruction in Section~\ref{sec.hollow} by changing $\fb$. Here, we employ $\fb=3$ and $30$, while omitting the case with $\fb=1$, since it does not give a good approximation as mentioned above. We tune $n_0$ and $\eB$ for each $\fb$ to adjust the peak time and peak flux of the forwardly synthesized light curves while keeping the other parameters the same as in Section~\ref{sec.hollow}.
Figure~\ref{fig.fb} shows the reconstructed jet structures and the corresponding light curves. Here, we adjusted $\log (n_0/\mathrm{cm}^{-3})=-2.95$ and $\log \eB=-3.61$ for $\fb=3$ while $\log (n_0/\mathrm{cm}^{-3})=-2.78$ and $\log \eB=-3.95$ for $\fb=30$. 
The initial inner truncation angle is $\Theta_0=0.172$ for $\fb=3$ while $\Theta_0=0.101$ for $\fb=30$. 
As shown in the left panel of Figure~\ref{fig.fb}, the reconstructed jet structures are qualitatively the same, irrespective of the value of $\fb$.  
Quantitatively speaking, the relative difference between the jet structures for $\fb=7$ and $\fb=30$ is relatively larger than that for $\fb=7$ and $\fb=3$, due to the larger $\Theta_0$ for $\fb=30$: 
In the case of $\fb=30$, a Gaussian shape is assumed for the wider edge part, $0.101 \le \theta \le \thj $, while a Gaussian shape is assumed for narrower part in the case of $\fb=7$, for which the jet shape is already different from the Gaussian at $\theta \sim 0.101$. The relative difference between $\fb=7$ and $\fb=30$ decreases as $\theta$ goes zero but remains at the level of $\gtrsim 30$ per cent.
We also note that the discrepancy between the used light curve for inversion and the forwardly synthesized one is smaller for larger $\fb$ while $\Tf$ becomes also smaller, as mentioned above, as shown in the right panel of Figure~\ref{fig.fb}.

\bsp	
\label{lastpage}
\end{document}